# Intracranial EEG structure-function coupling predicts surgical outcomes in focal epilepsy

Nishant Sinha[1,2], John S. Duncan[5,8], Beate Diehl[5], Fahmida A. Chowdhury[5], Jane de Tisi[5], Anna Miserocchi[5], Andrew W. McEvoy[5], Kathryn A. Davis[1,2], Sjoerd B. Vos[6,7,9], Gavin P. Winston[5,8,10], Yujiang Wang[3,4,5], Peter N. Taylor[3,4,5]

[1]Department of Neurology, Penn Epilepsy Center, Perelman School of Medicine, University of Pennsylvania, Philadelphia, PA, 19104, USA

[2]Center for Neuroengineering and Therapeutics, University of Pennsylvania, Philadelphia, PA, 19104, USA

[3]Translational and Clinical Research Institute, Faculty of Medical Sciences, Newcastle University, Newcastle upon Tyne, United Kingdom

[4]Computational Neuroscience, Neurology, and Psychiatry Lab (www.cnnp-lab.com), ICOS Group, School of Computing, Newcastle University, Newcastle upon Tyne, United Kingdom

[5]Department of Epilepsy, UCL Queen Square Institute of Neurology, London, WC1N 3BG, United Kingdom

[6]UCL Centre for Medical Image Computing, London, WC1V 6LJ, United Kingdom

[7]Neuroradiological Academic Unit, UCL Queen Square Institute of Neurology, London, WC1N 3BG, United Kingdom

[8] MRI Unit, Chalfont Centre for Epilepsy, Bucks, SL9 0RJ, United Kingdom

[9]Centre for Microscopy, Characterisation, and Analysis, The University of Western Australia, Nedlands, Australia

[10]Department of Medicine, Division of Neurology, Queen's University, Kingston, Canada

Correspondence to: Nishant Sinha
Address: Hayden Hall Room 301, University of Pennsylvania, 240 South 33rd Street, Philadelphia, PA 19104, USA
Email: nishant.sinha89@gmail.com
Twitter: @_Nishant_Sinha
Orcid ID: 0000-0002-2090-4889

Correspondence may also be addressed to: Peter Neal Taylor
Address: Urban Sciences Building, 1 Science Square, Newcastle Upon Tyne, NE4 5TG, Tyne and Wear, UK
Email: peter.taylor@newcastle.ac.uk
Orcid ID: 0000-0003-2144-9838

# Abstract

**Background:** Alterations to structural and functional brain networks have been reported across many neurological conditions. However, the relationship between structure and function—their coupling—is relatively unexplored, particularly in the context of an intervention. Epilepsy surgery alters the brain structure and networks to control the functional abnormality of seizures. Given that surgery is a structural modification aiming to alter the function, we hypothesized that stronger structure-function coupling, in the area to be resected, preoperatively is associated with a greater chance of post-operative seizure control.

**Method:** We constructed structural and functional brain networks in 39 subjects with medication-resistant focal epilepsy using multimodal data from intracranial EEG (iEEG) recordings (pre-surgery), structural MRI (pre-and post-surgery), and diffusion-weighted MRI (pre-surgery). We investigated pre-operative structure-function coupling at two spatial scales: a) at the global iEEG network level and b) at the resolution of individual iEEG electrode contacts using "virtual surgeries." By incorporating these structure-function coupling metrics and routine clinical variables in a cross-validated predictive model, we benchmarked their added value to predict seizure outcomes.

**Result:** At a global network level, seizure-free individuals had stronger structure-function coupling pre-operatively than those that were not seizure-free regardless of the choice of interictal segment or frequency band. At the resolution of individual iEEG contacts, the virtual surgery approach provided complementary information to localize epileptogenic tissues. In predicting seizure outcomes, structure-function coupling measures were more important than clinical attributes, and together they predicted seizure outcomes with an accuracy of 85% and sensitivity of 87%.

**Conclusion:** The underlying assumption that the structural changes induced by surgery translate to the functional level to control seizures is valid when the structure-functional coupling is strong. Mapping the regions that contribute to structure-functional coupling using virtual surgeries may help aid surgical planning.

# Introduction

Surgery is an effective therapy for many people with focal drug-resistant epilepsy[1,2]. Accurate localization and complete surgical removal of the epileptogenic zone are crucial for achieving seizure freedom[3]. Localizing the epileptogenic zone can sometimes be challenging with non-invasive methods alone. In this situation, intracranial electroencephalography (iEEG) may be employed, with electrodes implanted directly in contact with the cortex[4]. Unfortunately, even after iEEG implantation and surgery, 30-40% of subjects experience seizures in the short term and nearly 50% relapse in the long term[5–7]. To accurately inform the treatment plan and advise individuals before surgery, it is critical to assess how well the epileptogenic tissues are localized by iEEG and to identify those who are less likely to achieve seizure freedom by a planned resection[8].

Epileptogenic tissues remaining after surgery cause post-surgery seizure recurrence[9,10]. Reasons for incomplete resection of epileptogenic tissue include a) pre-surgical assessment did not fully localize this tissue, b) proximity to eloquent cortex precludes a complete resection that would have caused significant deficits, or c) a combination of both factors[9]. It is increasingly recognized that epileptogenic tissue constitutes a distributed network[11–14] rather than a well-circumscribed region[9]. Indeed, a circumscribed spatially contiguous resection in individuals with distributed epileptogenic tissues may not render them seizure-free if some epileptogenic network tissues are spared by surgery[15–18]. Thus, it would be beneficial to identify and quantify the epileptogenic tissue pre-operatively and measure the impact of a planned surgery on the epileptic network, to predict postoperative seizure outcomes[19,20].

Several studies have investigated structural and functional networks in epilepsy[19–33], but there is a lack of studies correlating the two modalities, particularly using iEEG[34–36]. Given that structural connectivity (SC) constrains functional connectivity (FC), and functional connectivity modulates the structural connectivity via plasticity mechanisms, these relationships may offer insights into mechanisms pertinent in intractable focal epilepsy[37–39]. A recent study showed increased coupling between structure and iEEG-derived functional networks by analyzing seizures in individuals with medication-resistant focal epilepsy[35]. The authors highlighted a structural substrate supporting seizure spread and suggested that complementary information in both modalities may help localize the brain regions producing seizures. Epilepsy surgery comprises a structural change that is aimed to control abnormal function. The efficacy of the intervention will depend on the structure-function relationships and how they are impacted by the surgery.

We hypothesized that surgical resection would control seizures when there is a strong coupling between brain structure and function. Our rationale is that a stronger structure-function coupling would be a more effective medium to relay the structural alterations in the epileptic network, caused by surgery, to the functional connectivity that is necessary for seizure expression. We tested our hypothesis on data from individuals with intractable epilepsy who had iEEG implantations to localize epileptogenic tissues to guide resective surgery. Initially, we quantified the coupling between structural and functional networks at a global iEEG network level. We next performed "virtual surgeries" to estimate the effect of removing individual brain areas on the remaining functional network as a spatial map. This mapping allowed us to translate the global-network metric of structure-function coupling to a regional metric that can elucidate how well the iEEG implantations captured the epileptogenic tissues. Finally, we highlight the importance of considering structure-function coupling measures alongside other pre-surgical clinical attributes of an individual to predict seizure outcomes after epilepsy surgery.

# Methods

## Participants

We studied 39 individuals with medication-resistant focal epilepsy recruited at the National Hospital of Neurology and Neurosurgery, London, UK. Each person underwent a comprehensive pre-surgical evaluation for localizing the epileptogenic tissue. We acquired MRI sequences with optimized epilepsy protocols and performed iEEG monitoring using a combination of grid, strip, and depth electrodes (electrocorticography, ECoG) or electrodes implanted stereotactically (sEEG). All data was reviewed in the multidisciplinary team meetings. Surgical resections were in the temporal lobe (n=22), frontal lobe (n=12), parietal lobe (n=3), and occipital lobe (n=2). After surgery, each individual was followed up for at least 12 months and were graded seizure outcomes using the ILAE classification of post-operative seizure outcome[40]. 15 individuals were seizure-free (ILAE 1) after surgery, and 24 had seizure recurrence (ILAE 2-5). Table 1 is a summary of clinical and demographics attributes, and Table S1 tabulates the entire data.

[Table 1]

The study was approved by the National Hospital for Neurology and Neurosurgery and the Institute of Neurology Joint Research Ethics Committee. We analysed the data in this study following Newcastle University Ethics Committee approval (reference 1804/2020).

## Magnetic resonance imaging and intracranial electrographic data acquisition

Pre-operatively, each person had anatomical T1-weighted MRI and diffusion-weighted MRI. X-ray CT was acquired after intracranial EEG electrode placement to localize electrode contacts. T1-weighted MRI was acquired 3-12 months after surgery to delineate the extent of resection.

MRI data were acquired on 3-T GE Signa HDx scanner (General Electric) with standard imaging gradients with a maximum strength of 40 $mTm^{-1}$ and a slew rate 150 $Tm^{-1}s^{-1}$. All data were acquired using a body coil for transmission and an eight-channel phased array coil for reception. Standard clinical sequences were performed including a coronal three-dimensional T1-weighted volumetric acquisition (matrix = 256×256×170; in-plane resolution = 0.9375×0.9375 mm, slice thickness = 1.1 mm).

Diffusion MRI data was acquired using a cardiac-triggered single-shot spin-echo planar imaging sequence with echo time=73ms. Sets of 60 contiguous 2.4mm thick axial slices were obtained covering the whole brain, with diffusion sensitizing gradients applied in each of 52 noncollinear directions (b-value = 1200$mm^2s^{-1}$, $\delta$ = 21ms, $\Delta$ = 29ms using full gradient strength of 40$mTm^{-1}$) along with 6 non-diffusion weighted scans. The field of view was 24×24cm, and the acquisition matrix size was 96×96, zero-filled to 128×128 during reconstruction, giving a reconstructed voxel size of 1.875×1.875×2.4mm. The diffusion MRI acquisition time for a total of 3480 image slices was approximately 25min.

Intracranial EEG was sampled at 512Hz or 1024Hz. Depending on the clinical necessity, ECoG or SEEG were implanted to localize the epileptogenic tissue. Table 1 shows the type of implantation and number of implanted electrodes in each person. We extracted 1-hour segments of interictal EEG data at least 2 hours away from seizures, as identified by the clinical team.

We acquired post-implantation whole-brain CT images with 171 contiguous, 1mm thick axial slices with a matrix size of 512x512 and voxel size 0.43×0.43×1mm to map the implanted electrodes.

## Data pre-processing

We linearly registered the postoperative T1-weighted MRI to preoperative T1-weighted MRI using the FSL-FLIRT algorithm[41]. We manually drew a resection mask for every subject, ensuring high inter-rater agreement to identify surgically resected tissue[20]. We ran the FreeSurfer 'recon-all' pipeline on preoperative T1-weighted MRI to generate grey and white matter surfaces.

Diffusion MRI data were corrected for signal drift, followed by eddy current and movement artefacts correction using the FSL eddy_correct tool[42], and b-vectors were rotated using the FSL fdt-rotate-bvecs tool. We applied generalized q-sampling imaging (GQI) reconstruction in DSI studio with diffusion sampling length ratio of 1.25 followed by deterministic tractography[43]. Tractography generated approximately 2,000,000 tracts per person with tracking parameters configured as follows: Runge-Kutta method with step size 1mm, whole-brain seeding, initial propagation direction set to all fibre orientations, minimum tract length 15mm, maximum tract length 300mm, and topology informed pruning applied with one iteration to remove false connections. Linear registration converted the tracts generated in diffusion space to preoperative T1-weighted MRI space.

We processed iEEG data in three steps, as in our previous study[22]: a) removing artefactual channels by visual inspection, b) applying common average reference to all remaining channels, and c) filtering each channel with a notch filter at 50 Hz and 100 Hz (infinite impulse response filter with Q factor = 50, 4th order zero-phase lag) and bandpass filter (Butterworth 4th order zero-phase lag) between 1 and 70 Hz. For frequency-band specific analysis, we separately filtered the interictal signals in six frequency bands—delta (1-4 Hz), theta (4-8 Hz), alpha (8-13 Hz), beta (13-30 Hz), gamma (30-80 Hz), and high gamma (80-150 Hz).

To delineate electrode contacts overlapping with the tissue that was subsequently resected, we linearly registered the CT image to pre-surgery T1-weighted MRI space and marked the electrode coordinates semi-automatically[22]. We deemed any electrode within 5 mm of the surgically resected tissue to be resected and marked the other contacts as spared contacts. Figure 1(a-g) illustrates an overview of different imaging modalities and outputs after each data pre-processing step.

[Figure 1]

## Network generation

For each individual, we estimated the structural and functional connectivity networks between brain areas implanted by electrodes. The brain tissue underlying each electrode comprised the nodes of the network; thus, the number of network nodes equalled the number of implanted electrodes (excluding those affected by artefacts).

Estimation of structural connectivity required delineating tracts intercepted by each electrode. We applied the following—atlas agnostic—steps to delineate these tracts from the whole brain tractography data in Figure 1g. First, for each contact in the neocortical grey matter (pial) surface, we found the corresponding coordinate on the white matter surface. For contacts in the white matter, or deep brain structures (e.g., hippocampus), this step was not necessary. Second, we sampled all tracts that passed within a spherical diameter of 2mm, connecting at least two electrodes. Third, we computed the total number of tracts between electrodes to measure structural connectivity between electrodes. Since the spherical diameters of all network nodes are identical this connectivity metric can be considered equivalent to streamline density, as used in a wide range of other studies[28,44]. Figure 1(h-j) illustrates the tracts between contacts and structural connectivity network in one case. To verify the robustness of our results, we recomputed the structural connectivity networks with alternative diffusion metric capturing average tract length between contacts, and by varying spherical diameter at each contact with a different threshold at 5mm (Figure S1, S2).

We estimated functional connectivity between each electrode pair from the pre-processed, one-hour interictal segments of electrophysiological data filtered in broad-band (1-70Hz) and six frequency bands[22]. We applied Pearson correlation to 2-second sliding windows (without overlap) and averaged the correlation matrices over all windows to obtain one functional connectivity network. Figure 1(k, l) illustrates the functional connectivity network in one case, highlighting the structurally connected (direct) and structurally unconnected (indirect) functional connections.

Figure S7 illustrates the reproducibility of two widely reported findings in our network data[45]: a) significantly stronger mean functional connectivity between structurally connected node pairs compared to structurally unconnected node pairs, and b) decline in the strength of functional connectivity between structurally connected and unconnected node pairs in relation to Euclidean distance.

## Structure-function coupling analysis and virtual resection approach

We modelled the relationship between structural and functional networks in an epileptic brain at a global iEEG network level resolution. We computed the non-parametric Spearman's rank correlation between the connections present in the structural network and the corresponding connections in the functional network. Figure 2(a, b) illustrates the conceptualization of structure-function coupling in one case before surgery.

To investigate the potential impact of surgery on structure-function coupling at a more fine-grained resolution, we performed "virtual surgeries" on brain networks, illustrated in Figure 2(c-e). First, we removed a node and its corresponding connections in the structural and functional connectivity networks. Second, we recomputed Spearman's rank correlation between the structural and functional connections of the remaining network. Third, we obtained the difference between the structure-function coupling before surgery and after node removal. This difference reflected the impact of removing that node on the structure-function coupling of the remaining network[35,46]. We repeated these steps by removing one node at a time, noting the changes in the structure-function coupling each time, and ultimately expressing these changes as z-score in each subject's brain network.

[Figure 2]

Virtual resections quantified coupling changes at a spatial resolution of an individual node by estimating the effect of removing individual brain areas on the remaining network. Performing virtual surgeries across all network nodes allowed us to generate a spatial map that quantified expected structural-function coupling changes at a regional level. Figure 2f shows a spatial map for one case illustrating the structure-function coupling changes expected by removing individual regions. We quantified the localizing value of these spatial maps by measuring the effect size of discrimination between the resected and spared tissues by computing the area under the receiver operating characteristic curve, AUC. AUC is a patient-specific, non-parametric summary statistic that verifies associations between SC-FC coupling changes and seizure outcomes after surgery.

We also applied another variation of the virtual resection approach to ensure robustness[23,48]. Instead of removing one node at a time, we removed all nodes deemed resected, as defined by the postoperative MRI, and compared the change in SC-FC coupling with the changes due to random resection of the same number of nodes; Figure S4 describes this approach in detail.

### Predictive model design

We implemented a support vector machine (SVM) learning algorithm to assess the generalisability of coupling measures when used alongside other clinical attributes to predict patient-specific chances of seizure freedom after surgery[15,49]. Table 1 summarises the 15 features—13 clinical variables and 2 coupling measures—incorporated as the SVM's input. Binary seizure outcomes—labelled as seizure-free or not seizure-free—were the SVM's output. We implemented a linear kernel in SVM because this enabled a direct interpretation of weights as relative feature importance. We applied a leave-one-out cross-validation scheme in which the data was divided into training and testing folds. SVMs were trained on the training fold and tested on the data in the test fold, thus enabling assessment of the out-of-sample generalization.

We applied SVM recursive feature elimination strategy to assess feature importance. Specifically, in the first iteration of SVM training and testing, we included all 15 features. We recorded the performance of SVM (AUC ± 95% CI) and ranked the 15 features in the order of their relative feature importance. Ranking features indicated metrics that were important for prediction performance as opposed to those that may be confounding. In the second iteration of SVM training and testing, we removed the least important feature and reassessed SVM performance with 14 features. We continued eliminating the least important features recursively until only a single feature remained, thus, identifying the combination of most informative features that maximized generalization of seizure outcome prediction in our cohort.

### Statistical analysis

To model the coupling between structure and function, we performed robust estimation to obtain the regression line and computed non-parametric Spearman's rank correlation. These approaches are less sensitive to the effect of outliers and appropriate when the normal distribution in data cannot be assumed.

For testing the hypothesis that the structure-function coupling is higher in seizure-free individuals than those that are not seizure-free, we applied a one-tailed non-parametric Wilcoxon rank sum test and effect size as Cohen's $d$ score. Results were declared significant for $p<0.05$. To correct multiple comparisons, we applied Benjamini-Hochberg false discovery rate correction at a significance level of 5%. In simulations of virtual resections, we computed the effect size of discrimination between spared and resected tissue non-parametrically by calculating the area under the receiver operating characteristic curve (AUC). In machine learning analysis for assessing generalisability, we report 95% confidence intervals of AUC using a bias-corrected and accelerated percentile method from 10,000 bootstraps resamples with replacement.

### Data availability

We make available all the anonymized brain networks, electrode coordinates with resection indicators, and clinical attributes of 39 subjects included in this study at https://doi.org/zenodo.

# Results

The main objective of our study was to assess the coupling between structural and functional networks derived from diffusion-weighted MRI and iEEG. First, we investigated if stronger structure-function coupling before surgery is an indicator of seizure-free surgical outcomes at a global iEEG network level. Second, we mapped the estimated impact of removing individual brain areas on increase/decrease in structure-function coupling to quantify how well the iEEG implantations captured the epileptogenic tissues. Third, we assessed the potential predictive value of using structure-function coupling measures during the pre-surgical workup, alongside common clinical attributes, to determine post-surgery seizure outcomes.

Table 1 summarises an individual's clinical attributes and highlights that these attributes, including the type of electrode implantations (SEEG vs. ECoG) or number of implanted electrode contacts amongst others, did not differentiate between seizure outcomes.

## Stronger structure-function coupling before surgery in individuals that are seizure-free after surgery

Brain areas implanted by iEEG electrodes had stronger coupling between structural and functional networks in seizure-free subjects than those that were not seizure-free. Figure 3(a-f) illustrates iEEG implantations in two cases: case 1 was not seizure-free and case 2 was seizure-free after surgery. At the location of iEEG implantation, Figure 3(b, e) maps the patient-specific structural and interictal functional connectivity matrices measured from the brain regions underlying each electrode. We quantified the SC-FC coupling by computing the non-parametric Spearman's ranked correlation between the connections present in the SC network and the corresponding connections in the FC network. We found that the seizure-free case 2 had a higher SC-FC coupling (rho = 0.36 [95%CI 0.28, 0.44]) than the not seizure-free case 1 (rho = 0.20 [95%CI 0.12, 0.26]).

**[Figure 3]**

Across the entire cohort, we detected significantly higher SC-FC coupling in the seizure-free group than the not seizure-free group (p = 0.002, d = 0.76). Figure 3(g) illustrates the consistency of our findings for three different choices of interictal time segments to estimate functional connectivity. Figure 3(h) shows the frequency band-limited estimation of interictal functional connectivity to compute SC-FC coupling. We did not find that a specific frequency band drives the significantly higher SC-FC coupling in seizure-free individuals, rather it was consistent across the frequency bands.

We estimated variations of structural connectivity by changing the size of the regions modelled by the sphere centred at the location of each electrode coordinate and using alternative diffusion MRI metric. Figure S1 and Figure S2 illustrates the consistency of our results to these variations in estimating

SC. We also investigated if simple geometric measures such as the Euclidian distance between the implanted electrode contacts explain surgical outcomes. Structural connectivity, functional connectivity, and Euclidian distance between contacts capture complementary but different organizational properties of the brain network. Figure S3 demonstrates that SC-FC coupling explains the association with surgical outcomes more strongly than Euclidian distance. Figure S6 shows the effect of implant and SC-FC coupling for each group with ILAE seizure outcomes at one-year after surgery.

In summary, surgery—a structural procedure—is more effective in controlling the abnormal functional dynamics—seizures—when it is performed on brain networks with stronger coupling between structural and functional connectivity.

### Epileptogenic tissues are localized by coupling boosters

Next, we investigated how well the iEEG implantations captured epileptogenic tissue. We mapped the estimated impact of removing individual brain areas on increase/decrease in structure-function coupling. We applied a virtual resection approach at the network node (region) level to quantify these coupling changes. Specifically, we removed individual nodes (one at a time), recomputed the SC-FC coupling of the remaining network, and quantified the change from the baseline SC-FC coupling of the full network. Quantifying these coupling changes enabled us to compute a spatial map at the resolution of individual brain areas that estimated the impact of removing that brain area on the network level structure-function relationships. By assessing seizure outcomes and overlap of surgery with the spatial maps of the estimated SC-FC coupling changes, we analysed how good the localization of the epileptogenic tissues was.

**[Figure 4]**

Figure 4(a, b) colour codes the electrode contacts of the same two cases based on the SC-FC coupling changes. Red contacts with lower negative z-scores are coupling boosters—removing these contacts boosted the coupling of the expected remaining network. Blue contacts with higher positive z-scores are coupling dampers—removing these contacts dampened the coupling of the expected remaining network. The lower horizontal panel in Figure 4(a, b) plots the electrode contacts sorted by SC-FC coupling changes. The electrode contacts marked in black in the panel above the colour bar indicates the resected electrode that overlapped with the surgery mask. We found that surgery overlapped more with coupling dampers in not seizure-free individuals, whereas in those that were seizure-free, surgery overlapped more with coupling boosters. These results indicate that the iEEG electrodes identified as coupling boosters can localize epileptogenic tissues.

With each electrode characterized in each individual as coupling booster/damper, we computed non-parametric effect size (AUC) for discriminating resected and spared contacts. This effect size is a patient-specific measure that quantifies a) the probability of resecting coupling dampers for $0.5<AUC\leq1$,

b) probability of resecting coupling boosters for 0<AUC<0.5, and c) chance level probability of resecting coupling boosters or dampers for AUC=0.5. This AUC measure is identical to distinguishability statistics[22]. Figure 4(c, e) illustrate the discrimination between resected and spared electrodes in two cases and Figure 4d plots the effect size between seizure-free and not seizure-free groups. We found that the probability of resecting coupling dampers in individuals that are not seizure free is significantly higher than the probability of resecting coupling boosters in those that are seizure free ($p = 0.007$, $d = 0.96$ [95% CI 0.34, 1.56]). Therefore, the discrimination between resected and spared contacts characterized as coupling boosters and dampers is an important metric to determine seizure outcomes.

We verified the consistency of our findings with a different virtual resection strategy in Figure S4. In this alternative approach, we removed all electrodes deemed resected by surgery, recomputed the SC-FC coupling of the remaining network, and statistically quantified the change in SC-FC coupling against 1000 instances of randomly removing the same number of electrodes. In Figure S5, we illustrate consistent results from an MRI derived expected post-surgery structural network. In summary, our results consistently show that surgery is more associated with seizure-free outcomes if it resects those brain areas whose removal is expected to boost the structure-function coupling of the remaining network.

## Structure-function coupling measurements complement clinical variables in predicting seizure outcomes

In previous sections, we measured the structure-function relationships at the two spatial scales: a) at the resolution of the entire iEEG network comprised of brain areas implanted with electrodes, and b) at the resolution of individual brain areas, estimating their impact on network level structure-function relationships. In this section, we treated these novel measures quantifying structure-function relationships as bivariate features. We assessed if these measures contained complementary information to discriminate surgical outcomes more effectively. Figure 5(a) plots each subject on a 2D plane defined by the two coupling measures and draws linear decision boundaries for mapping the likelihood of seizure recurrence. We found that these coupling measures were not correlated ($r = -0.21$, $p = 0.19$); and combining them in a bivariate model discriminated seizure-free and not seizure-free individuals with the area under receiver operator characteristic curve (AUC) of 0.79.

**[Figure 5]**

Next, we combined the two coupling measures with 13 common clinical attributes to predict seizure outcomes. We incorporated these 15 measures as input features in a linear SVM model and implemented a leave-one-out cross-validation scheme to predict seizure recurrence. We noted the AUC for predicting seizure recurrence and the relative importance of input features in making that prediction. We removed the least important feature and re-assessed the performance of the SVM model in predicting seizure recurrence. This recursive feature elimination technique enabled the detection of the best combination of features that led to the highest AUC. Figure 5b illustrates each step of recursive feature elimination, with each feature ranked in the order of their relative importance. AUC was maximum at 0.83 [95%CI 0.70, 0.95] when SVM combined coupling metrics, the two most important features, with six clinical variables: number of anti-seizure medications (ASMs) taken before surgery, lesional vs non-lesional MRI, number of electrodes implanted, temporal lobe epilepsy (TLE) vs extra temporal lobe epilepsy (ETLE), epilepsy onset age, and history of status epilepticus. Figure 5c shows the ROC curve with the confusion matrix drawn at the optimal operating point. These results indicate that our measures quantifying structure-function relationships are important factors that can be used alongside routine clinical variables to predict post-surgery seizure outcomes.

# Discussion

We analysed the structure-function coupling in individuals using diffusion-weighted MRI and iEEG. We found that stronger coupling between structure and function was associated with seizure-free outcomes after surgery. Virtual resection of individual brain areas revealed that individuals have a higher likelihood of seizure freedom if the structure-function coupling was expected to increase after the surgery. By mapping cortical tissues underlying each iEEG contact as coupling boosters or dampers, we show that the structure-function relationships can be probed at the resolution of individual brain areas. This can be done during iEEG monitoring to improve the depiction of the extent of epileptogenic tissues. The coupling between structure and function is an important predictor of seizure outcome that could be used alongside clinical variables to identify individuals who are less likely to achieve seizure freedom by a planned resection.

Strong coupling between the brain's structural and functional network at rest is a hallmark of a healthy brain[38,45,50,51]; in comparison, a drop in SC-FC coupling is associated with epilepsy and longer epilepsy duration[52]. In healthy subjects, numerous studies suggested moderate to tight coupling; a recent study demonstrated that the structure-function coupling could be as high as $0.9 \pm 0.1$ at group level and $0.55 \pm 0.1$ at individual subject level[50]. In people with epilepsy, Zhang et al. demonstrated reduced SC-FC coupling, with the greatest reductions in those with longest disease duration[52]. We found that in individuals with medication-resistant focal epilepsy, there is a weaker structure-function coupling in those who are not seizure-free after surgery compared to those who are seizure free. Our results indicate that refractory epilepsy is likely associated with disruption of structure-function relationships, and when there is a substantial loss of coupling strength, surgery is less likely to control seizures.

In subjects implanted with iEEG electrodes, it is crucial to determine if the epileptogenic zone is sampled entirely and if an individual has a high chance of seizure remission[19,53]. Using only the iEEG data, previous work developed quantitative measures including node strength, seizure likelihood[23,54], brain network ictogenicity[32], source-sink score[55], neural fragility[56], synchronizability index[57] amongst others, to address these critical needs. Despite surgical resection being a structural procedure to control function, most previous work overlooked the structure-function relationships of brain areas implanted by iEEG electrodes[35,58,59]. By measuring the patient-specific structural brain network sampled by iEEG electrodes, in this study, we mapped the expected impact of resecting brain tissues underlying each electrode on changes in structure-function coupling. Our results suggest that removing epileptogenic cortex may cause the structure-function coupling of the remaining network to become stronger, thus normalizing patients towards controls. However, if surgery does not remove the epileptogenic zone, there is failure to control seizures and structure-function coupling of the remaining network can weaken. We have shown the expected impact of removing individual brain areas on structure-function coupling strength as a spatial map using only the pre-surgical data. We envision these spatial maps could help probe the overlap of a planned resection with epileptogenic tissues and the likelihood of seizure freedom with high accuracy during an individual's pre-surgical monitoring.

What is the added value in identifying structure-function relationships during pre-surgical workup over other measures already available in clinical practice? While some clinical variables like presence of focal to bilateral tonic-clonic seizures are indicators of poor seizure outcomes, most studies report inconsistent findings; features found predictive of seizure outcome in some studies are not predictive in others[60]. The analysis by Bell et al. is one the most comprehensive studies that combined 27 clinical variables on a mixed cohort of people with temporal and extratemporal epilepsy to estimate the probability of seizure freedom[49]. Jehi et al. incorporated some of these variables on nomograms to evaluate the risk of poor surgical outcomes[61]. Our recent work benchmarked non-invasive surgical outcome measures against 13 clinical variables, and other studies adopted a similar approach to benchmarking novel quantitative measures from iEEG network analysis against those already proposed[19]. Such comparisons demonstrate the added value of new measures. We envision combining these multimodal measures with unimodal measures derived from structural or functional imaging modalities and clinical data to make a comprehensive software tool for accurately identifying individuals who are less likely to achieve seizure freedom.

Our study has limitations and important caveats that future studies should address. First, we could not directly perform a case-control analysis to compare structure-function coupling between controls and individuals. People without epilepsy rarely undergo iEEG implantations, and it is difficult to estimate the strength of structure-function coupling in a healthy population in areas where individuals had iEEG implantations. However, future studies can leverage our recently proposed normative iEEG atlas as a promising alternative to circumvent the challenges associated with case-control analysis with iEEG data[62,63]. Second, our analysis does not focus on reducing the invasiveness of iEEG. iEEG is amongst the most invasive diagnostic tools and should be used sparingly on carefully selected individuals who are highly likely to benefit from undergoing iEEG related surgical procedures. Future studies could expand our analysis to a multilayer framework for making iEEG minimally invasive or replacing it with complementary non-invasive modalities (e.g., fMRI, MEG, or high-density scalp EEG)[64]. Third, our analysis does not investigate the networks that remain after surgery. Estimating the expected remaining network follows our previous approaches, designed for future prospective applications for any intended surgery. Nonetheless, validation on actual post-surgery data is important, and it can highlight the mechanisms of network changes due to surgery and its relation to outcomes[19,20]. Finally, our analysis only incorporated the interictal/seizure-free epochs of iEEG data, which could be both a strength and a weakness. Many studies reported remarkable state changes in the pre-ictal and ictal epochs that we do not analyse in this study[35]. However, a major strength is that making predictions from seizure-free epochs can substantially reduce the time a person remains in an epilepsy monitoring unit with iEEG electrodes implanted waiting for multiple seizure episodes.

In conclusion, we have shown that SC-FC coupling is an important measure that is related to seizure freedom after surgery. Structural alteration by surgery is more likely to control the abnormal functional dynamics associated with seizures when the coupling between structure and function is high. We suggest that mapping the impact of coupling changes at the resolution of individual brain areas can better evaluate surgical outcomes and create choices for alternative resection strategies, thus assisting the planning of epilepsy surgeries.

# Acknowledgements

NS acknowledges funding from NINDS R01NS116504 and Research Excellence Academy, Newcastle University, UK. PNT was supported by the Wellcome Trust (105617/Z/14/Z and 210109/Z/18/Z) and UKRI Future Leaders Fellowship (MR/T04294X/1). YW was supported by the Wellcome Trust (208940/Z/17/Z) and UKRI Future Leaders Fellowship (MR/V026569/1). KD acknowledges funding from NINDS R01NS116504. JD and SBV were funded by the UCLH NIHR BRC. Scan acquisition and GPW were supported by the MRC (G0802012, MR/M00841X/1). We acknowledge support from the Epilepsy Society. This work was supported by the National Institute for Health Research University College London Hospitals Biomedical Research Centre. The authors acknowledge the facilities and scientific and technical assistance of the National Imaging Facility, a National Collaborative Research Infrastructure Strategy (NCRIS) capability, at the Centre for Microscopy, Characterisation, and Analysis, the University of Western Australia.

# Competing interests

The authors report no competing interests.

# References

1. Brodie MJ, Barry SJE, Bamagous GA, Norrie JD, Kwan P. Patterns of treatment response in newly diagnosed epilepsy. *Neurology*. 2012;78(20):1548-1554. doi:10.1212/wnl.0b013e3182563b19

2. Devinsky O, Vezzani A, O'Brien TJ, et al. Epilepsy. *Nat Rev Dis Primers*. 2018;4(1):18024. doi:10.1038/nrdp.2018.24

3. Zijlmans M, Zweiphenning W, Klink N van. Changing concepts in presurgical assessment for epilepsy surgery. *Nat Rev Neurol*. 2019;15(10):594-606. doi:10.1038/s41582-019-0224-y

4. Duncan JS, Winston GP, Koepp MJ, Ourselin S. Brain imaging in the assessment for epilepsy surgery. *Lancet Neurology*. 2016;15(4):420-433. doi:10.1016/s1474-4422(15)00383-x

5. Tisi J de, Bell GS, Peacock JL, et al. The long-term outcome of adult epilepsy surgery, patterns of seizure remission, and relapse: a cohort study. *Lancet Lond Engl*. 2011;378(9800):1388-1395. doi:10.1016/s0140-6736(11)60890-8

6. Jobst BC, Cascino GD. Resective Epilepsy Surgery for Drug-Resistant Focal Epilepsy: A Review. *Jama*. 2015;313(3):285. doi:10.1001/jama.2014.17426

7. Téllez-Zenteno JF, Wiebe S. Long-term seizure and psychosocial outcomes of epilepsy surgery. *Curr Treat Option Ne*. 2008;10(4):253-259. doi:10.1007/s11940-008-0028-7

8. Spencer S, Huh L. Outcomes of epilepsy surgery in adults and children. *Lancet Neurology*. 2008;7(6):525-537. doi:10.1016/s1474-4422(08)70109-1

9. Rosenow F, Lüders H. Presurgical evaluation of epilepsy. *Brain*. 2001;124(9):1683-1700. doi:10.1093/brain/124.9.1683

10. Vakharia VN, Duncan JS, Witt J, Elger CE, Staba R, Engel J. Getting the best outcomes from epilepsy surgery. *Ann Neurol*. 2018;83(4):676-690. doi:10.1002/ana.25205

11. Bartolomei F, Lagarde S, Wendling F, et al. Defining epileptogenic networks: Contribution of SEEG and signal analysis. *Epilepsia*. 2017;58(7):1131-1147. doi:10.1111/epi.13791

12. Jehi L. The Epileptogenic Zone: Concept and Definition. *Epilepsy Curr*. 2018;18(1):12-16. doi:10.5698/1535-7597.18.1.12

13. Spencer SS. Neural Networks in Human Epilepsy: Evidence of and Implications for Treatment. *Epilepsia*. 2002;43(3):219-227. doi:10.1046/j.1528-1157.2002.26901.x

14. Kramer MA, Cash SS. Epilepsy as a Disorder of Cortical Network Organization. *Neurosci*. 2012;18(4):360-372. doi:10.1177/1073858411422754

15. Sinha N, Davis KA. Mapping Epileptogenic Tissues in MRI-Negative Focal Epilepsy. *Neurology*. 2021;97(16):754-755. doi:10.1212/wnl.0000000000012696

16. Keller SS, Glenn GR, Weber B, et al. Preoperative automated fibre quantification predicts postoperative seizure outcome in temporal lobe epilepsy. *Brain J Neurology*. 2016;140(Pt 1):68-82. doi:10.1093/brain/aww280

17. Galovic M, Baudracco I, Wright-Goff E, et al. Association of Piriform Cortex Resection With Surgical Outcomes in Patients With Temporal Lobe Epilepsy. *Jama Neurol*. 2019;76(6):690. doi:10.1001/jamaneurol.2019.0204

18. Concha L, Kim H, Bernasconi A, Bernhardt BC, Bernasconi N. Spatial patterns of water diffusion along white matter tracts in temporal lobe epilepsy. *Neurology*. 2012;79(5):455-462. doi:10.1212/wnl.0b013e31826170b6

19. Sinha N, Wang Y, Silva NM da, et al. Structural brain network abnormalities and the probability of seizure recurrence after epilepsy surgery. *Neurology*. 2021;96(5):10.1212/WNL.0000000000011315. doi:10.1212/wnl.0000000000011315

20. Taylor PN, Sinha N, Wang Y, et al. The impact of epilepsy surgery on the structural connectome and its relation to outcome. *Neuroimage Clin*. 2018;18:202-214. doi:10.1016/j.nicl.2018.01.028

21. Sinha N, Peternell N, Schroeder GM, et al. Focal to bilateral tonic–clonic seizures are associated with widespread network abnormality in temporal lobe epilepsy. *Epilepsia*. 2021;62(3):729-741. doi:10.1111/epi.16819

22. Wang Y, Sinha N, Schroeder GM, et al. Interictal intracranial electroencephalography for predicting surgical success: The importance of space and time. *Epilepsia*. 2020;(January):epi.16580. doi:10.1111/epi.16580

23. Sinha N, Dauwels J, Kaiser M, et al. Predicting neurosurgical outcomes in focal epilepsy patients using computational modelling. *Brain*. 2016;140(2):319-332. doi:10.1093/brain/aww299

24. Munsell BC, Gleichgerrcht E, Hofesmann E, et al. Personalized connectome fingerprints: Their importance in cognition from childhood to adult years. *Neuroimage*. 2020;221:117122. doi:10.1016/j.neuroimage.2020.117122

25. Gleichgerrcht E, Keller SS, Drane DL, et al. Temporal Lobe Epilepsy Surgical Outcomes Can Be Inferred Based on Structural Connectome Hubs: A Machine Learning Study. *Ann Neurol*. 2020;88(5):970-983. doi:10.1002/ana.25888

26. Bonilha L, Jensen JH, Baker N, et al. The brain connectome as a personalized biomarker of seizure outcomes after temporal lobectomy. *Neurology*. 2015;84(18):1846-1853. doi:10.1212/wnl.0000000000001548

27. Englot DJ, D'Haese PF, Konrad PE, et al. Functional connectivity disturbances of the ascending reticular activating system in temporal lobe epilepsy. *J Neurology Neurosurg Psychiatry*. 2017;88(11):925. doi:10.1136/jnnp-2017-315732

28. Morgan VL, Rogers BP, Anderson AW, Landman BA, Englot DJ. Divergent network properties that predict early surgical failure versus late recurrence in temporal lobe epilepsy. *J Neurosurg*. 2019;132(5):1324-1333. doi:10.3171/2019.1.jns182875

29. Morgan VL, Rogers BP, González HFJ, Goodale SE, Englot DJ. Characterization of postsurgical functional connectivity changes in temporal lobe epilepsy. *J Neurosurg*. 2020;133(2):392-402. doi:10.3171/2019.3.jns19350

30. Goodale SE, González HFJ, Johnson GW, et al. Resting-State SEEG May Help Localize Epileptogenic Brain Regions. *Neurosurgery*. 2019;86(6):792-801. doi:10.1093/neuros/nyz351

31. Lagarde S, Roehri N, Lambert I, et al. Interictal stereotactic-EEG functional connectivity in refractory focal epilepsies. *Brain J Neurology*. 2018;141(10):2966-2980. doi:10.1093/brain/awy214

32. Goodfellow M, Rummel C, Abela E, Richardson MP, Schindler K, Terry JR. Estimation of brain network ictogenicity predicts outcome from epilepsy surgery. *Sci Rep-uk*. 2016;6(1):29215. doi:10.1038/srep29215

33. Sinha N, Wang Y, Dauwels J, et al. Computer modelling of connectivity change suggests epileptogenesis mechanisms in idiopathic generalised epilepsy. *Neuroimage Clin*. 2019;21:101655. doi:10.1016/j.nicl.2019.101655

34. Ashourvan A, Shah P, Pines A, et al. Pairwise maximum entropy model explains the role of white matter structure in shaping emergent co-activation states. *Commun Biology*. 2021;4(1):210. doi:10.1038/s42003-021-01700-6

35. Shah P, Ashourvan A, Mikhail F, et al. Characterizing the role of the structural connectome in seizure dynamics. *Brain*. 2019;142(7):1955-1972. doi:10.1093/brain/awz125

36. Carr SJA, Gershon A, Shafiabadi N, Lhatoo SD, Tatsuoka C, Sahoo SS. An Integrative Approach to Study Structural and Functional Network Connectivity in Epilepsy Using Imaging and Signal Data. *Frontiers Integr Neurosci*. 2021;14:491403. doi:10.3389/fnint.2020.491403

37. Chu CJ, Tanaka N, Diaz J, et al. EEG functional connectivity is partially predicted by underlying white matter connectivity. *Neuroimage*. 2014;108:23-33. doi:10.1016/j.neuroimage.2014.12.033

38. Honey CJ, Sporns O, Cammoun L, et al. Predicting human resting-state functional connectivity from structural connectivity. *Proc National Acad Sci*. 2009;106(6):2035-2040. doi:10.1073/pnas.0811168106

39. Hermundstad AM, Bassett DS, Brown KS, et al. Structural foundations of resting-state and task-based functional connectivity in the human brain. *P Natl Acad Sci Usa*. 2013;110(15):6169-6174. doi:10.1073/pnas.1219562110

40. Wieser HG, Blume WT, Fish D, et al. Proposal for a New Classification of Outcome with Respect to Epileptic Seizures Following Epilepsy Surgery. *Epilepsia*. 2001;42(2):282-286. doi:10.1046/j.1528-1157.2001.35100.x

41. Jenkinson M, Beckmann CF, Behrens TEJ, Woolrich MW, Smith SM. FSL. *Neuroimage*. 2012;62(2):782-790. doi:10.1016/j.neuroimage.2011.09.015

42. Andersson JLR, Sotiropoulos SN. An integrated approach to correction for off-resonance effects and subject movement in diffusion MR imaging. *Neuroimage*. 2016;125:1063-1078. doi:10.1016/j.neuroimage.2015.10.019

43. Yeh FC, Wedeen VJ, Tseng WYI. Estimation of fiber orientation and spin density distribution by diffusion deconvolution. *Neuroimage*. 2011;55(3):1054-1062. doi:10.1016/j.neuroimage.2010.11.087

44. Bonilha L, Nesland T, Martz GU, et al. Medial temporal lobe epilepsy is associated with neuronal fibre loss and paradoxical increase in structural connectivity of limbic structures. *J Neurology Neurosurg Psychiatry*. 2012;83(9):903-909. doi:10.1136/jnnp-2012-302476

45. Goni J, Heuvel MP van den, Avena-Koenigsberger A, et al. Resting-brain functional connectivity predicted by analytic measures of network communication. *Proc National Acad Sci*. 2013;111(2):833-838. doi:10.1073/pnas.1315529111

46. Alstott J, Breakspear M, Hagmann P, Cammoun L, Sporns O. Modeling the impact of lesions in the human brain. *Plos Comput Biol*. 2009;5(6):e1000408. doi:10.1371/journal.pcbi.1000408

47. Conrad EC, Bernabei JM, Kini LG, et al. The sensitivity of network statistics to incomplete electrode sampling on intracranial EEG. *Netw Neurosci Camb Mass*. 2020;4(2):484-506. doi:10.1162/netn_a_00131

48. Sinha N, Dauwels J, Wang Y, Cash SS, Taylor PN. An in silico approach for pre-surgical evaluation of an epileptic cortex. *2014 36th Annu Int Conf Ieee Eng Medicine Biology Soc*. 2014;2014:4884-4887. doi:10.1109/embc.2014.6944718

49. Bell GS, Tisi J de, Gonzalez-Fraile JC, et al. Factors affecting seizure outcome after epilepsy surgery: an observational series. *J Neurology Neurosurg Psychiatry*. 2017;88(11):933-940. doi:10.1136/jnnp-2017-316211

50. Sarwar T, Tian Y, Yeo BTT, Ramamohanarao K, Zalesky A. Structure-function coupling in the human connectome: A machine learning approach. *Neuroimage*. 2021;226:117609. doi:10.1016/j.neuroimage.2020.117609

51. Suárez LE, Markello RD, Betzel RF, Misic B. Linking Structure and Function in Macroscale Brain Networks. *Trends Cogn Sci*. 2020;24(4):302-315. doi:10.1016/j.tics.2020.01.008

52. Zhang Z, Liao W, Chen H, et al. Altered functional–structural coupling of large-scale brain networks in idiopathic generalized epilepsy. *Brain*. 2011;134(10):2912-2928. doi:10.1093/brain/awr223

53. Morgan VL, Johnson GW, Cai LY, et al. MRI network progression in mesial temporal lobe epilepsy related to healthy brain architecture. *Netw Neurosci*. 2021;Just Accepted(Just Accepted):1-36. doi:10.1162/netn_a_00184

54. Sinha N, Dauwels J, Kaiser M, et al. Reply: Computer models to inform epilepsy surgery strategies: prediction of postoperative outcome. *Brain*. 2017;140(5):e31-e31. doi:10.1093/brain/awx068

55. Gunnarsdottir KM, Li A, Smith RJ, et al. Source-sink connectivity: A novel interictal EEG marker for seizure localization. *Biorxiv*. Published online 2021:2021.10.15.464594. doi:10.1101/2021.10.15.464594

56. Li A, Huynh C, Fitzgerald Z, et al. Neural fragility as an EEG marker of the seizure onset zone. *Nat Neurosci*. Published online 2021:1-10. doi:10.1038/s41593-021-00901-w

57. Proix T, Bartolomei F, Guye M, Jirsa VK. Individual brain structure and modelling predict seizure propagation. *Brain J Neurology*. 2017;140(3):641-654. doi:10.1093/brain/awx004

58. Betzel RF, Medaglia JD, Kahn AE, Soffer J, Schonhaut DR, Bassett DS. Structural, geometric and genetic factors predict interregional brain connectivity patterns probed by electrocorticography. *Nat Biomed Eng*. 2019;3(11):902-916. doi:10.1038/s41551-019-0404-5

59. Chiang S, Stern JM, Engel J, Haneef Z. Structural–functional coupling changes in temporal lobe epilepsy. *Brain Res*. 2015;1616:45-57. doi:10.1016/j.brainres.2015.04.052

60. Bonilha L, Keller SS. Quantitative MRI in refractory temporal lobe epilepsy: relationship with surgical outcomes. *Quantitative Imaging Medicine Surg*. 2015;5(2):204-224. doi:10.3978/j.issn.2223-4292.2015.01.01

61. Jehi L, Yardi R, Chagin K, et al. Development and validation of nomograms to provide individualised predictions of seizure outcomes after epilepsy surgery: a retrospective analysis. *Lancet Neurology*. 2015;14(3):283-290. doi:10.1016/s1474-4422(14)70325-4

62. Taylor PN, Papasavvas CA, Owen TW, et al. Normative brain mapping of interictal intracranial EEG to localise epileptogenic tissue. *Brain*, 2022. doi:10.1093/brain/awab380

63. Bernabei J, Sinha N, Campbell AT, et al. Normative intracranial EEG maps epileptogenic tissues in focal epilepsy. *Brain*. 2022. doi:10.1093/brain/awab480

64. Ramaraju S, Wang Y, Sinha N, et al. Removal of Interictal MEG-Derived Network Hubs Is Associated With Postoperative Seizure Freedom. *Front Neurol*. 2020;11:563847. doi:10.3389/fneur.2020.563847

**Table 1: Demographic and clinical data**

| Variables, n \ Groups | Seizure Free | Not Seizure Free | Significance |
|---|---|---|---|
| Subjects | 15 | 24 | |
| Sex (male/female) | 8/7 | 8/16 | $\chi^2 = 0.81$<br>$p = 0.37$ |
| Side of surgery (n, Left/Right) | 8/7 | 12/12 | $\chi^2 = 0.02$<br>$p = 0.90$ |
| Hippocampal sclerosis n (%) | 4 (26.7%) | 3 (12.5%) | $\chi^2 = 0.48$<br>$p = 0.49$ |
| Preoperative MRI (n, normal/abnormal) | 4/11 | 12/12 | $\chi^2 = 1.22$<br>$p = 0.27$ |
| History of FBTCS n (%) | 9 (60%) | 16 (66.7%) | $\chi^2 = 0.01$<br>$p = 0.94$ |
| History of status epilepticus n (%) | 4 (26.7%) | 5 (20.8%) | $\chi^2 < 0.01$<br>$p = 0.98$ |
| Electrode implantation type (ECoG/sEEG) | 4/11 | 7/17 | $\chi^2 = 0.04$<br>$p = 0.84$ |
| Surgery location (n, TLE/eTLE) | 10/5 | 12/12 | $\chi^2 = 0.48$<br>$p = 0.49$ |
| Age at surgery (years, median ± IQR) | 28.2 ± 5.4 | 31.9 ± 17.4 | $d = 0.36$<br>$p = 0.46$ |
| Age at epilepsy onset (years, median ± IQR) | 10.0 ± 9.5 | 13.5 ± 11.5 | $d = 0.69$<br>$p = 0.07$ |
| Epilepsy duration (years, median ± IQR) | 22.5 ± 7.2 | 19.6 ± 10.4 | $d = 0.07$<br>$p = 0.57$ |
| All ASMs before surgery (n, median ± IQR) | 7 ± 4 | 7 ± 3 | $d = 0.07$<br>$p = 0.87$ |
| Total electrode contacts implanted (n, median ± IQR) | 73 ± 45 | 73 ± 34 | $d = 0.002$<br>$p = 0.97$ |

*Abbreviations*—IQR: interquartile range; TLE: temporal lobe epilepsy; eTLE: extratemporal lobe epilepsy; MRI: magnetic resonance imaging; ECoG: electrocorticography; sEEG: stereo-electroencephalography; FBTCS: focal to bilateral tonic-clonic seizures; ASM: antiseizure medication; n: sample size; p: two-tailed Wilcoxon rank-sum test; d: Cohen's d score for effect size.

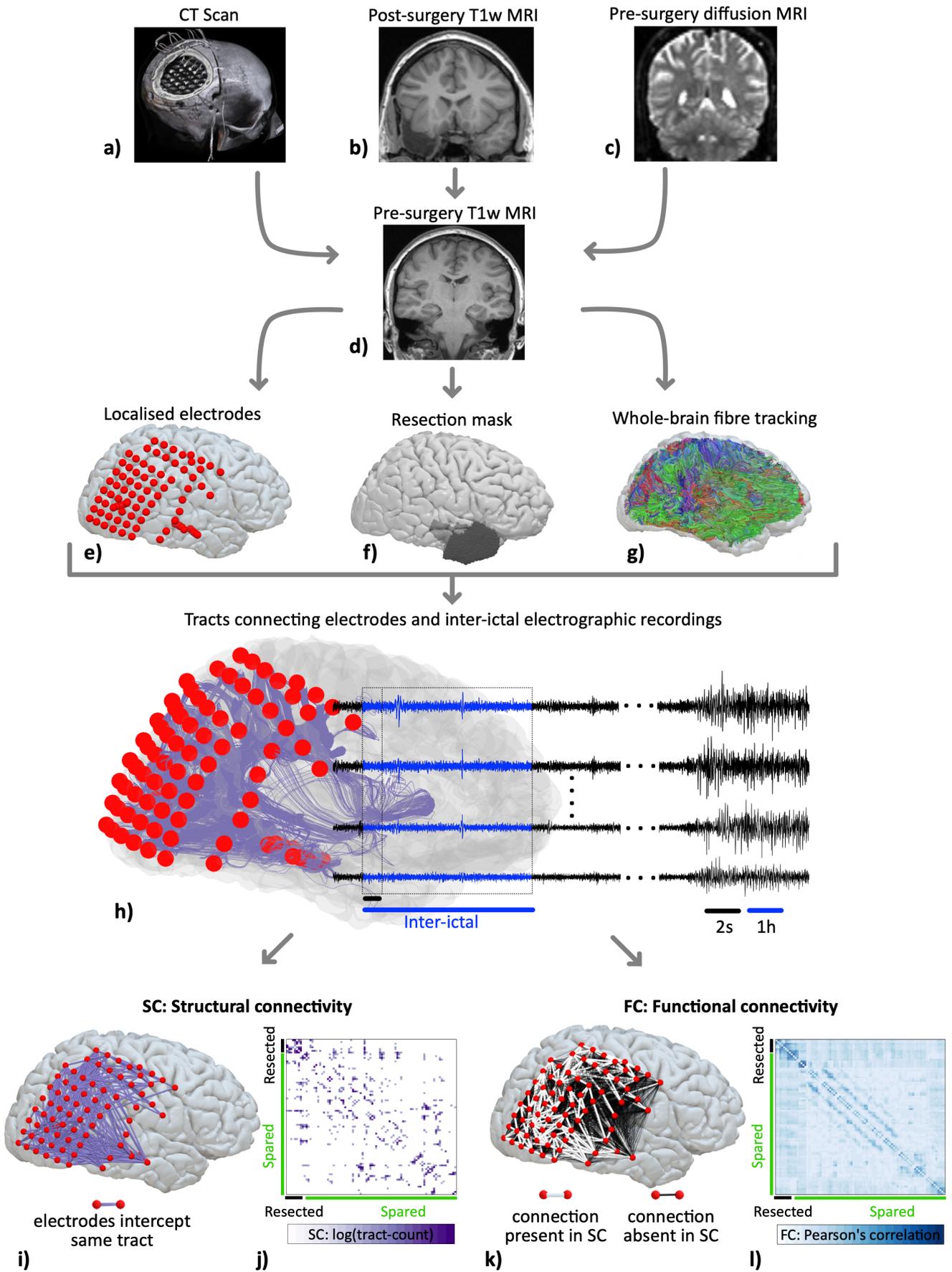

**Figure 1: Overview of network generation.** Panels **(a-d)** show different modalities acquired for each subject. We aligned the CT scan in panel **(a)** to the pre-surgery T1w MRI in panel **(d)** for delineating

the coordinates of implanted electrodes shown in panel **(e)**. We registered the post-surgery T1w MRI in **(b)** with the pre-surgery T1w MRI scan in **(a)** to manually draw a resection mask in the pre-surgery T1w MRI space illustrated in panel **(f)**. We performed the whole-brain fibre tracking on pre-surgery diffusion MRI in native space **(c)** and then aligned the tracts to the pre-surgery T1-MRI space shown in panel **(g)**. In panel **(h)** we combined electrode coordinates, tracts, and surgery information. The example case illustrated in the figure was not seizure-free after the surgery. From the whole-brain fibre tracts, we delineated the tracts connecting each electrode shown in purple. Each electrode (in red) records the electrophysiological signals (in black) directly from the cortical tissues. We analysed one-hour inter-ictal segments (in blue) at least two hours from seizures. By counting the number of tracts between each electrode, we constructed the structural connectivity. Panel **(i)** maps the binarized structural connectivity network for illustrating connections between electrodes. Panel **(j)** shows the weighted structural connectivity matrix with tract counts transformed on a log scale. Rows and columns of the connectivity matrix are the electrodes, spared electrodes labelled in green and resected electrodes labelled in black. Panel **(k)** depicts the functional connectivity network derived from 2-sec windows of one-hour interictal iEEG recordings. Functional connections with underlying structural connections are shown in white, and the remaining structurally unconnected functional connections are shown in black. Panel **(l)** shows the weighted functional connectivity matrix with electrodes in rows and columns reordered as spared (green) and resected (black) contacts.

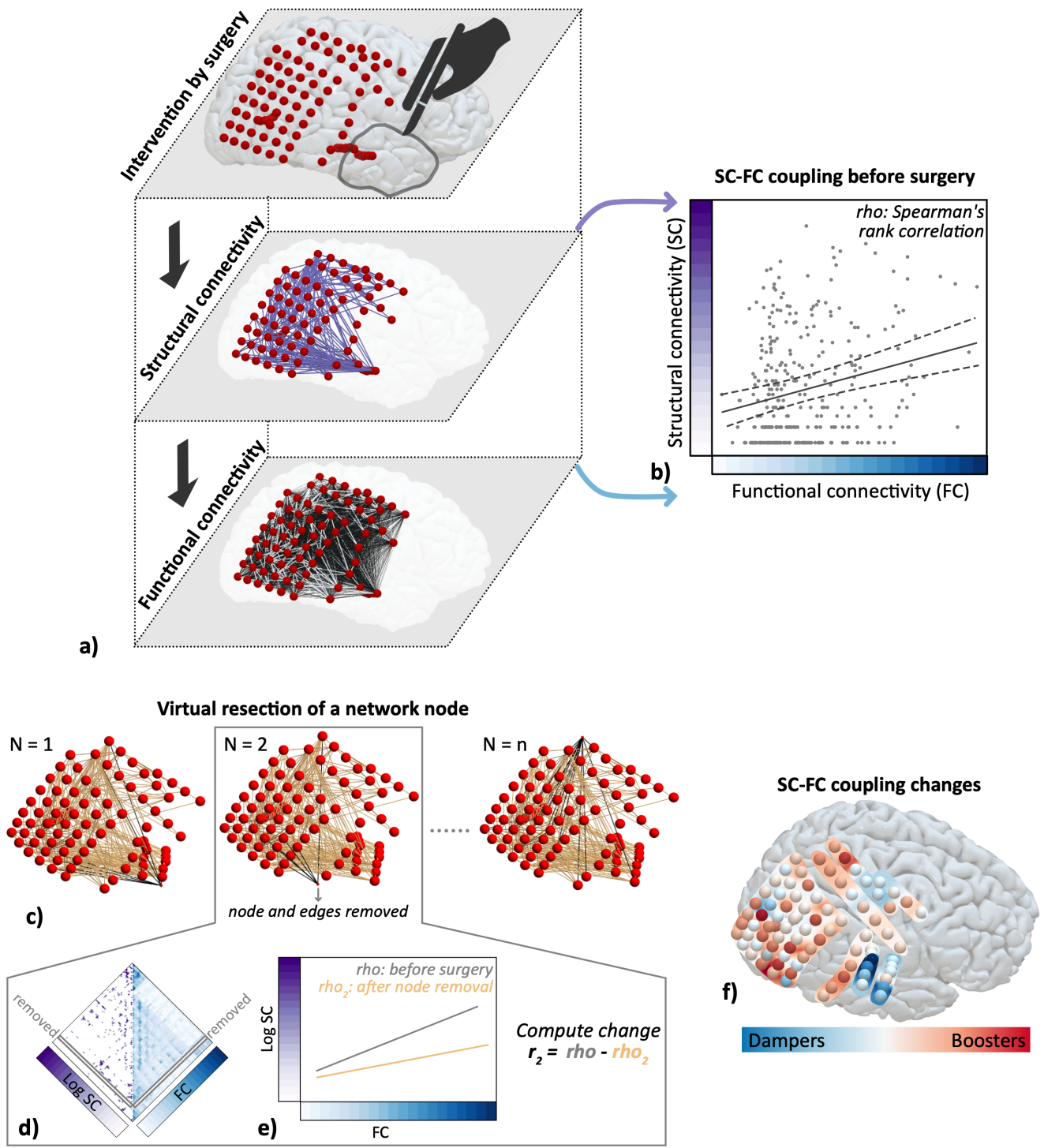

**Figure 2: Virtual resection approach for estimating changes in structure-function coupling.** Panel **(a)** conceptualizes a common framework to study surgical intervention and its impact on the structural connectivity (SC) and functional connectivity (FC) networks as three interlinked layers. Resecting localized cortical tissue by surgery alters the brain anatomy, and the underlying structural network. An assumption is that the structure and function are strongly interlinked and alteration to the structure would change the function to control the abnormal functional dynamics associated with seizures. Panel **(b)** illustrates structure-function coupling modelled by Spearman's rank correlation as a measure to evaluate the strength of interlinking between structure and function in brain areas sampled by iEEG electrodes.

Each point in the scatter plot represents a network connection. In panel **(c-e)**, we estimated the impact of removing individual brain areas on structure-function coupling by applying virtual resection of network nodes. Panel **(c)** illustrates the removal of a node (drawn smaller) and corresponding edges (in black). Removing a network node is equivalent to removing a row and a column from the structural and functional connectivity matrices in **(d)**. In panel **(e)**, we re-evaluated Spearman's rank correlation between the remaining connections of structure-function networks. We computed the change in coupling between networks after virtual resection and the original network with all nodes intact. Panel **(f)** maps the changes in SC-FC coupling. We detected that some cortical areas are coupling boosters (in red)—removing these nodes boosted the SC-FC coupling of the remaining network. Also, some other cortical areas are coupling dampers (in blue)—removing these nodes dampened the SC-FC coupling of the remaining network. Removing the red coupling boosters may improve the chance of seizure free outcomes; however, these areas are scattered indicating a distributed epileptogenic network. The illustrated case was not seizure-free after the surgery.

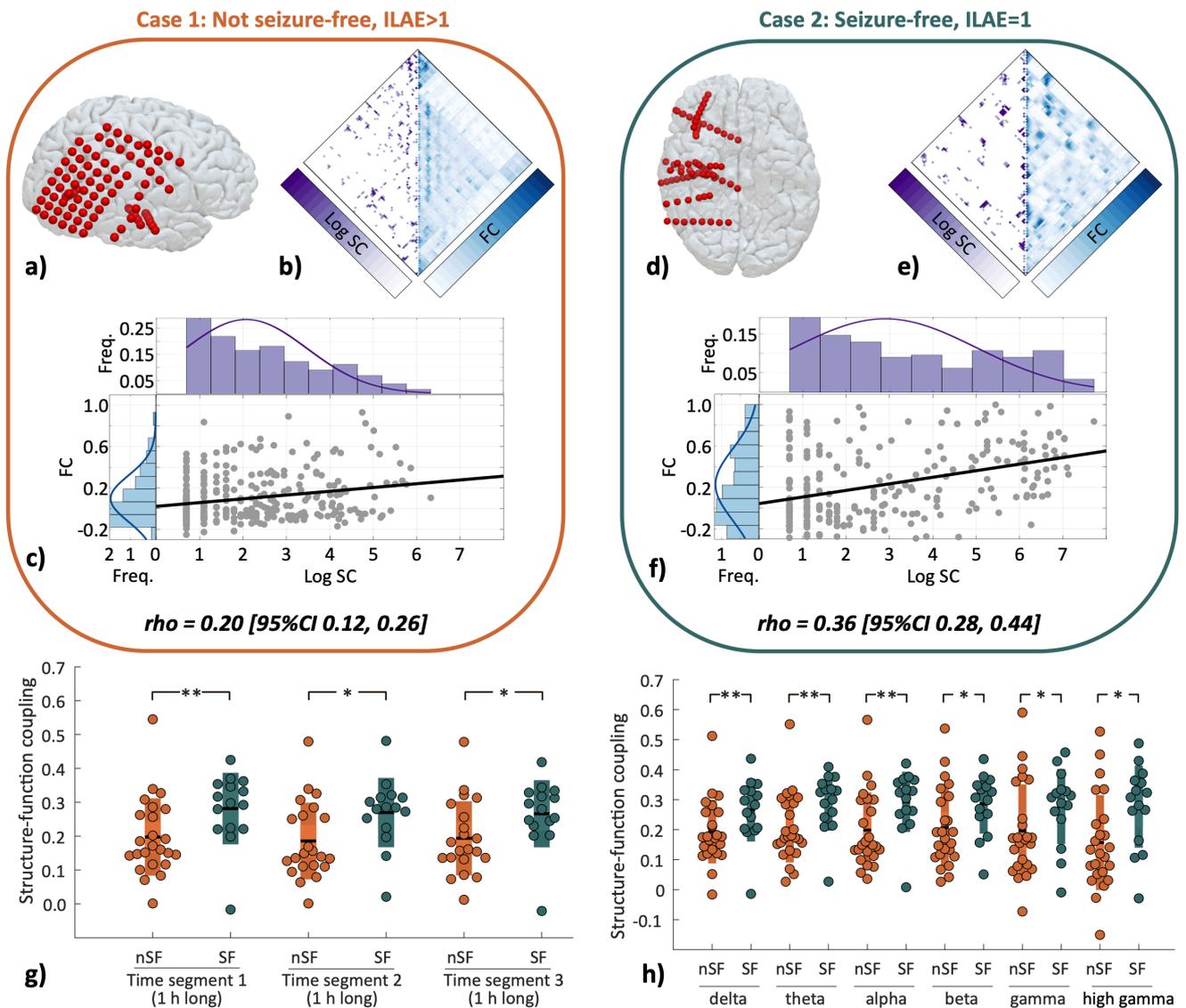

**Figure 3: Structure-function coupling is significantly higher in seizure-free group than in not seizure-free group.** Panel **(a-c)** illustrates example case 1 who was not seizure-free after surgery: electrode coordinates in **(a)**, structural connectivity (SC) and functional connectivity (FC) matrices in **(b)**, and SC-FC coupling between structural and functional edges in **(c)**. The histograms in **(c)** shows the distribution of structural and functional edges in purple and blue respectively. Panel **(d-f)** shows equivalent plots for an individual who is seizure-free after surgery. Seizure-free case 2 had significantly higher (non-overlapping 95% CI of rho) SC-FC coupling than the not seizure-free person 1. Panel **(g)** illustrates SC-FC coupling at group level between seizure-free (SF in teal) and not-seizure free (nSF in orange) individuals across three different interictal time segments of iEEG recordings. Those who were seizure free have significantly higher structure-function coupling than those who were not. *Statistical estimates. Segment 1: p = 0.002, d = 0.76; Segment 2: p = 0.009, d = 0.77; Segment 3: p = 0.008, d = 0.70*. In panel **(h)**, we evaluated structure-function coupling between seizure-free and not seizure-free individuals across different frequency bands. Across all frequency bands, the structure-function coupling is significantly higher in those that are seizure free than in those that are not. *Statistical estimates. delta*

*(1-4 Hz): p = 0.003, d = 0.68; theta (4-8 Hz): p = 0.001, d = 0.83; alpha (8-13 Hz): p = 0.003, d = 0.81; beta (13-30 Hz): p = 0.015, d = 0.66; gamma (30-80 Hz): p = 0.033, d = 0.56; and high gamma (80-150 Hz) p = 0.01, d = 0.79.*

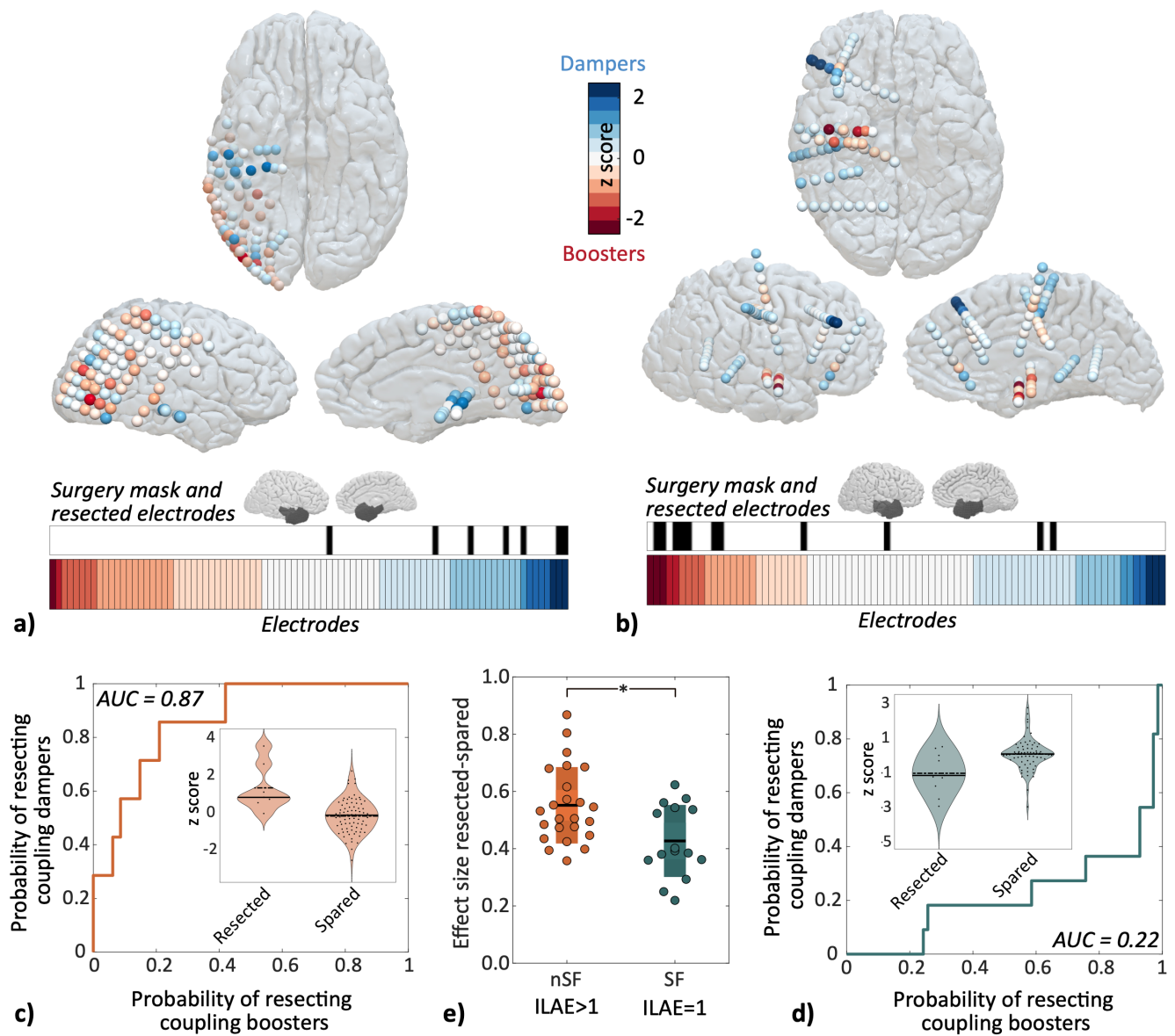

**Figure 4: Surgeries in not seizure-free individuals overlap with coupling dampers, whereas in those that are seizure free, surgery overlaps with coupling boosters.** Panel **(a)** plots the iEEG electrodes as coupling boosters and dampers for case 1 (not seizure-free). The surgery mask in sagittal view shows the location of the resected tissue. The colour bar plotted horizontally shows the implanted electrodes in case 1 sorted by coupling booster-damper metric and the binary plot in black highlights the resected electrodes. Panel **(b)** shows the equivalent contrasting plots for case 2 (seizure-free). Panel **(c)** quantifies the overlap between surgery and coupling booster-damper metric of electrodes for case 1 by computing the area under the receiver operating characteristic curve (AUC). Panel **(d)** shows the equivalent plot for case 2. AUC is a non-parametric effect size to discriminate between resected and spared tissues. This effect size is a patient-specific measure that quantifies the probability of resecting coupling dampers for $0.5<AUC\leq1$, resecting coupling boosters for $0<AUC<0.5$, chance level probability of resecting coupling boosters or dampers for $AUC=0.5$. The violin plots in the inset show coupling booster-damper data points for each electrode categorized as resected or spared. Panel **(e)** shows the

effect size between resected and spared tissues (AUC) is significantly higher in individuals that are not seizure-free than in those that are seizure free (p = 0.007, d = 0.96 [95% CI 0.34, 1.56]).

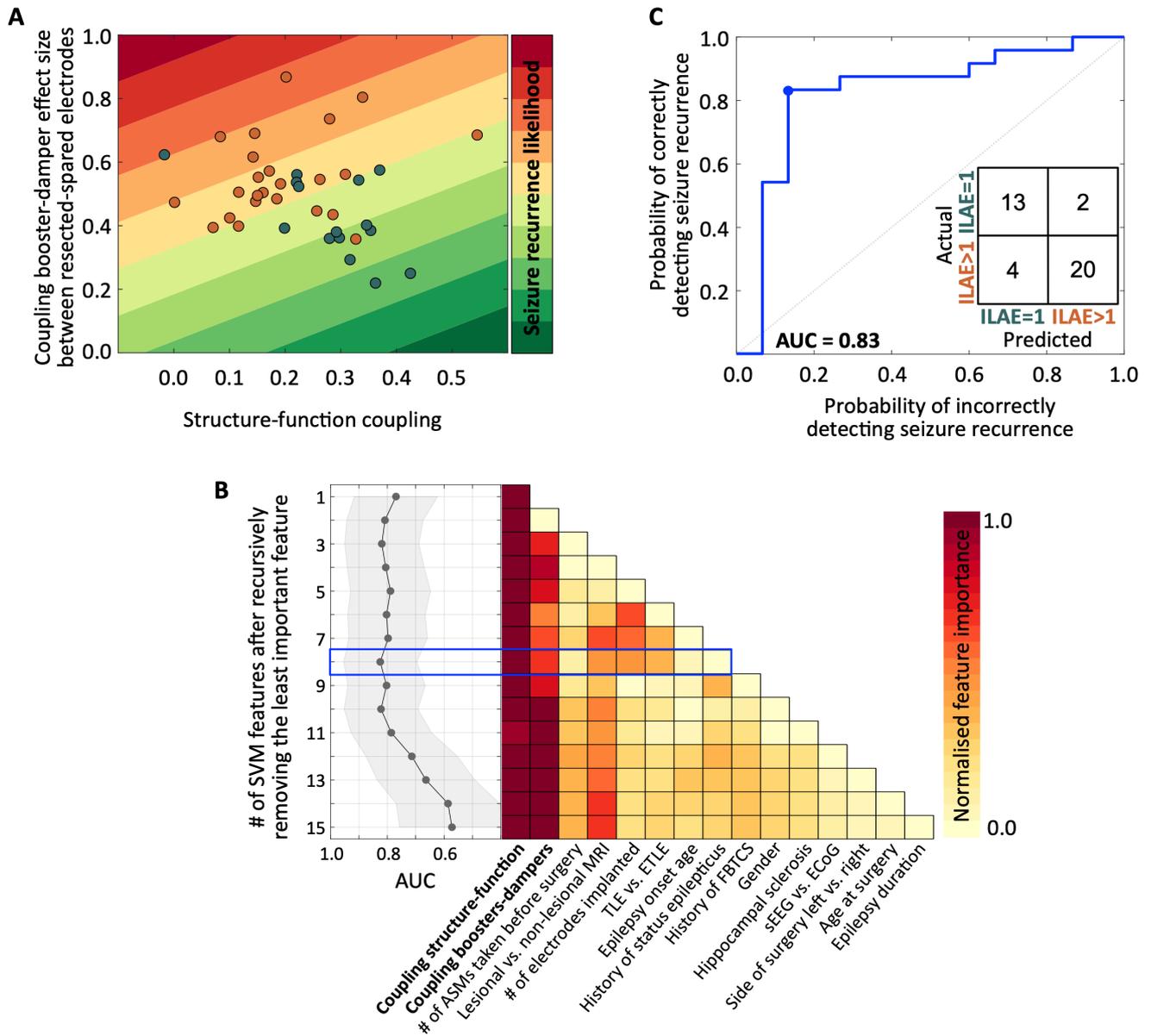

**Figure 5: Structure-function coupling measures rank high in feature importance compared to clinical variables.** Panel **(a)** shows the scatter plot between the structure-function coupling of brain networks before surgery and the discrimination between resected and spared tissue obtained from the coupling booster-damper metric. The two-dimensional plane is colour coded as a probability map of seizure recurrence likelihood. Each dot represents an individual; those that are seizure free are in teal, and those that are not seizure free are in orange. The two coupling measures are not correlated; together, they have complementary information that discriminates those that are seizure free from those that are not seizure free. **(b)** We incorporated the two coupling measures with 13 clinical features in the linear SVM model to predict seizure outcome after surgery. In the first iteration, we included all 15 features in the model, evaluated feature importance, and noted model performance. In the subsequent iterations, we recursively removed the least important feature and noted model performance until a single feature

remained. The colour plot maps the normalized feature importance after each round of recursive feature removal, and the graph on the left shows the AUC ± 95% CI. The blue box highlights the iteration at which AUC maximizes. Panel **(c)** shows the ROC curve at iteration 8 with the confusion matrix drawn in the inset. *Predictive performance estimates:* number of features = 8, AUC = 0.83 [95% CI 0.70, 0.95], at the optimal operating point (blue dot on ROC) accuracy = 85%, sensitivity = 87%, specificity = 83%.

# Supplementary: Intracranial EEG structure-function coupling predicts surgical outcomes in focal epilepsy

Nishant Sinha[1,2], John S. Duncan[5,8], Beate Diehl[5], Fahmida A. Chowdhury[5], Jane de Tisi[5], Anna Miserocchi[5], Andrew W. McEvoy[5], Kathryn A. Davis[1,2], Sjoerd B. Vos[6,7,9], Gavin P. Winston[5,8,10], Yujiang Wang[3,4,5], Peter N. Taylor[3,4,5]

[1]Department of Neurology, Penn Epilepsy Center, Perelman School of Medicine, University of Pennsylvania, Philadelphia, PA, 19104, USA

[2]Center for Neuroengineering and Therapeutics, University of Pennsylvania, Philadelphia, PA, 19104, USA

[3]Translational and Clinical Research Institute, Faculty of Medical Sciences, Newcastle University, Newcastle upon Tyne, United Kingdom

[4]Computational Neuroscience, Neurology, and Psychiatry Lab (www.cnnp-lab.com), ICOS Group, School of Computing, Newcastle University, Newcastle upon Tyne, United Kingdom

[5]Department of Epilepsy, UCL Queen Square Institute of Neurology, London, WC1N 3BG, United Kingdom

[6]UCL Centre for Medical Image Computing, London, WC1V 6LJ, United Kingdom

[7]Neuroradiological Academic Unit, UCL Queen Square Institute of Neurology, London, WC1N 3BG, United Kingdom

[8] MRI Unit, Chalfont Centre for Epilepsy, Bucks, SL9 0RJ, United Kingdom

[9]Centre for Microscopy, Characterisation, and Analysis, The University of Western Australia, Nedlands, Australia

[10]Department of Medicine, Division of Neurology, Queen's University, Kingston, Canada

Correspondence to: Nishant Sinha
Address: Hayden Hall Room 301, University of Pennsylvania, 240 South 33rd Street, Philadelphia, PA 19104, USA
Email: nishant.sinha89@gmail.com
Twitter: @_Nishant_Sinha
Orcid ID: 0000-0002-2090-4889

Correspondence may also be addressed to: Peter Neal Taylor
Address: Urban Sciences Building, 1 Science Square, Newcastle Upon Tyne, NE4 5TG, Tyne and Wear, UK
Email: peter.taylor@newcastle.ac.uk
Orcid ID: 0000-0003-2144-9838

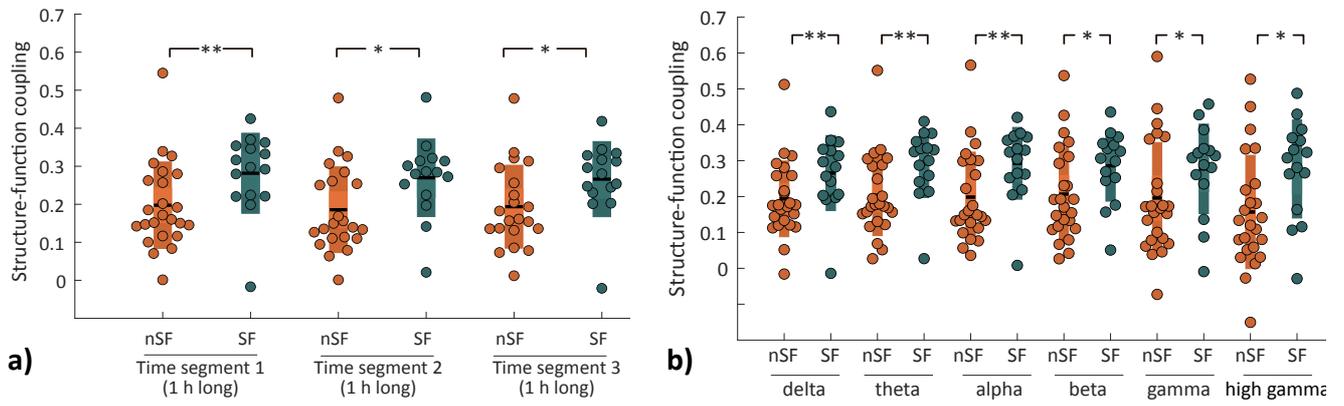

*Figure S1: Consistent results with structural connectivity networks estimated by applying a sphere diameter of 5mm on each iEEG contacts.* As opposed to the main manuscript where we estimated the structural connectivity between iEEG contacts by placing a sphere of 2mm diameter at each contact, here we estimated structural connectivity by applying an alternative threshold by placing a sphere of 5mm diameter at each contact. Consistent results are shown. Panel **(a)** illustrates SC-FC coupling at group level between seizure-free (SF in teal) and not seizure-free (nSF in orange) individuals across three different interictal time segments of iEEG recordings. Seizure-free individuals have significantly stronger structure-function coupling than not seizure-free individuals. *Segment 1: p = 0.002, d = 0.76; Segment 2: p = 0.009, d = 0.77; Segment 3: p = 0.008, d = 0.69*. In panel **(b)**, we evaluated structure-function coupling between seizure-free and not seizure-free individuals across different frequency bands. Across all frequency bands, the structure-function coupling is significantly higher in seizure-free individuals than in not seizure-free individuals. *Statistical estimates. delta (1-4 Hz): p = 0.004, d = 0.68; theta (4-8 Hz): p < 0.001, d = 0.83; alpha (8-13 Hz): p = 0.003, d = 0.81; beta (13-30 Hz): p = 0.015, d = 0.66; gamma (30-80 Hz): p = 0.03, d = 0.56; and high gamma (80-150 Hz) p = 0.01, d = 0.79.*

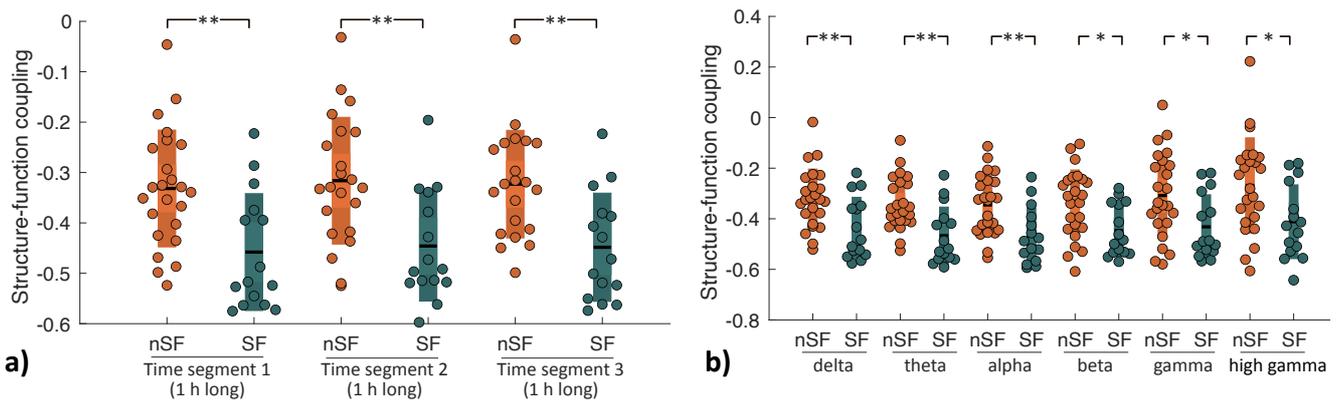

*Figure S2: Consistent results with structural connectivity networks estimated from tract length between iEEG electrode contacts.* We estimated the structural connectivity between iEEG contacts from the mean tract length metric of the diffusion MRI tractography data as opposed to tract counts metric shown in the main manuscript. Tract count and tract length metrics are inversely related. As expected, the coupling between structure-function with structure networks inferred from mean tract length, are in the opposite direction compared to those shown in manuscript. Panel **(a)** illustrates SC-FC coupling at group level between seizure-free (SF in teal) and not seizure-free (nSF in orange) individuals across three different interictal time segments of iEEG recordings. Seizure-free individuals have significantly stronger structure-function coupling than not seizure-free individuals. *Statistical estimates. Segment 1: p = 0.001, d = 1.1; Segment 2: p = 0.002, d = 1.1; Segment 3: p = 0.002, d = 1.2.* In panel **(b)**, we evaluated structure-function coupling between seizure-free and not seizure-free individuals across different frequency bands. Across all frequency bands, the structure-function coupling is significantly higher in seizure-free individuals than in not seizure-free individuals. *Statistical estimates. delta (1-4 Hz): p = 0.004, d = 0.98; theta (4-8 Hz): p < 0.001, d = 1.16; alpha (8-13 Hz): p = 0.001, d = 1.06; beta (13-30 Hz): p = 0.007, d = 0.84; gamma (30-80 Hz): p = 0.014, d = 0.81; and high gamma (80-150 Hz) p = 0.007, d = 0.83.*

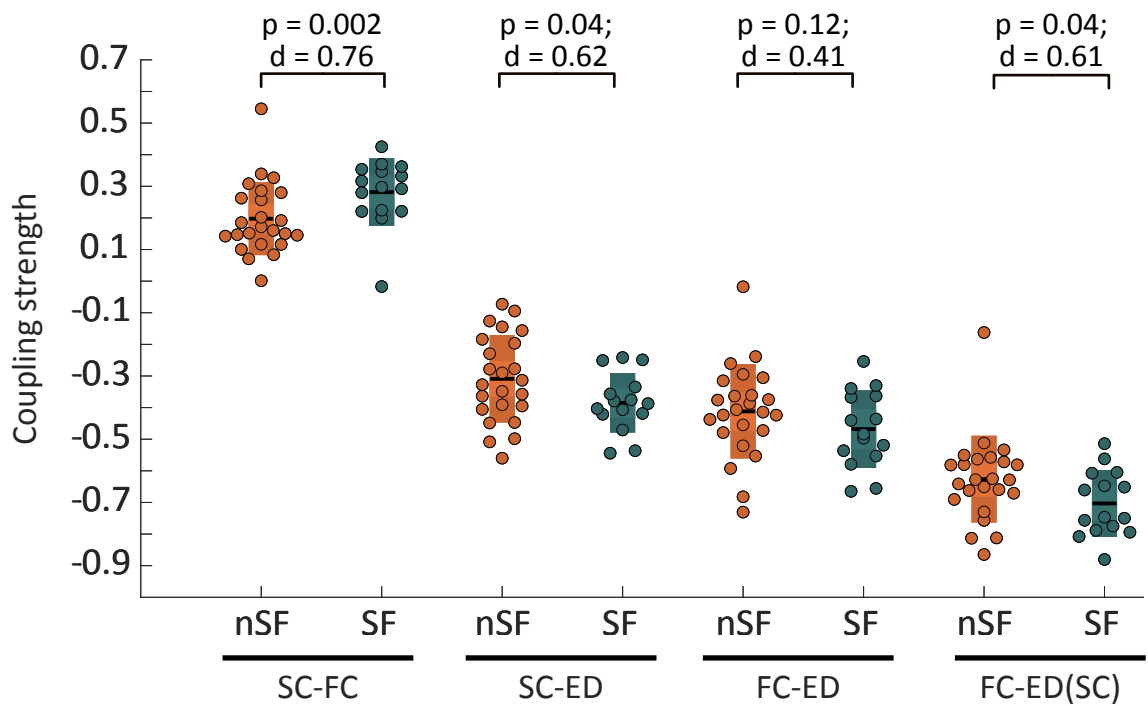

*Figure S3: Structure-function coupling is associated with surgical outcomes more strongly than coupling of structure or function with Euclidian distance between iEEG contacts.* The box plot shows the comparison between coupling strength from a) structural and functional connectivity between iEEG contacts (SC-FC), b) structural connectivity and Euclidean distance between contacts (SC-ED), c) functional connectivity and Euclidean distance between contacts (FC-ED), d) functional connectivity and Euclidean distance between contacts which are connected by a direct structural connections (FC-ED(SC)). For each measure of coupling strength, the effect size distinguishing the seizure-free (SF) and not seizure-free individuals (nSF) are shown. Compared to other measures involving Euclidian distance, structure-function coupling strength discriminated seizure-free and not seizure-free individuals with the highest effect size of *d = 0.76*.

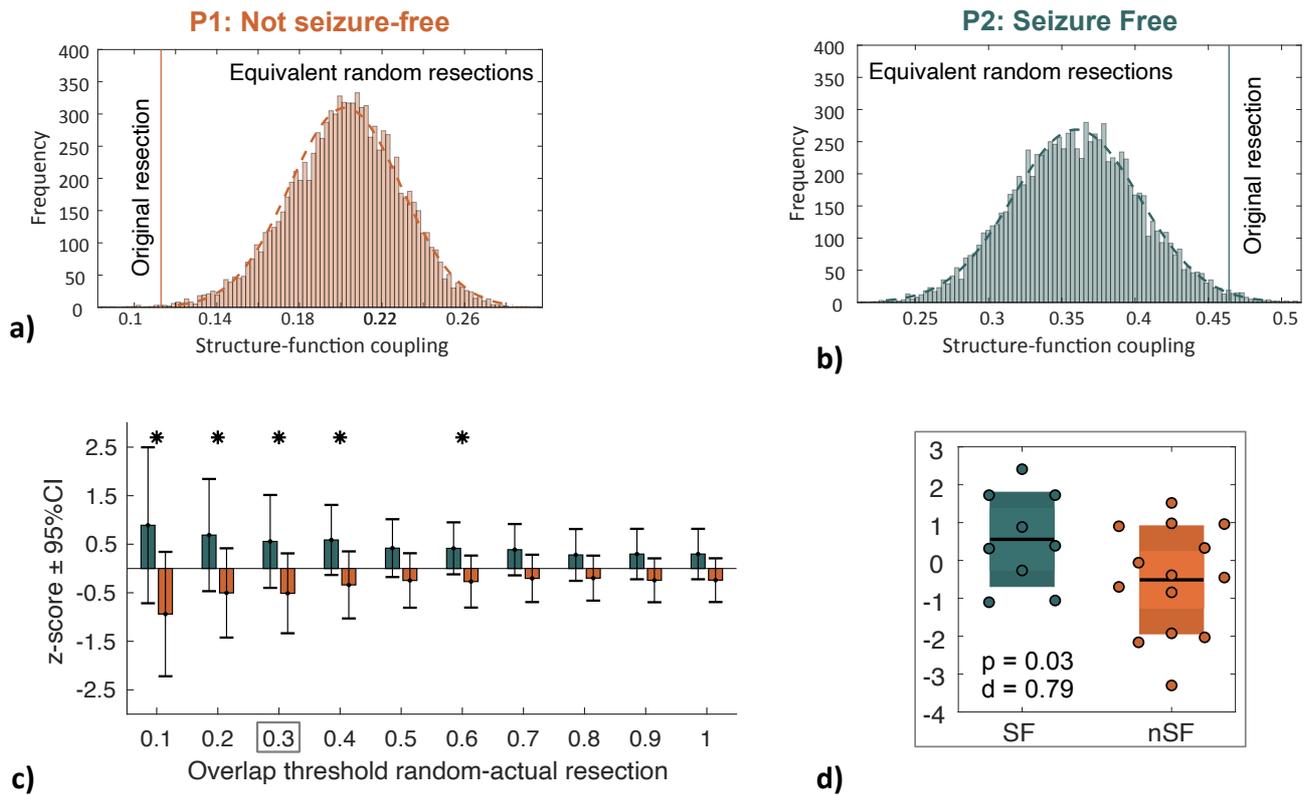

*Figure S4: Results are robust with an alternative virtual resection analysis method. In the main manuscript, we performed virtual surgeries by removing one node at a time.* An alternative could be to remove a set of nodes removed during actual resection and compare the change in structure-function coupling with the same number of nodes removed randomly (following the approach in *Sinha et al. 2017, Brain*)[23]. In panels a) and b), we have shown the same two cased as in the main manuscript. Consistent with our results in the main manuscript, we find that in panel a) for the not seizure-free case, the structure-function coupling was expected to reduce compared to random resections. In panel b) for the seizure-free case, the structure-function coupling was expected to reduce compared to random resections. A caveat of this approach is that when a more extensive resection removed more contacts, only a few contacts would remain to compare the random surgeries unless the random resections overlap with the actual resection. Thus, to expand the results to the entire cohort, we performed a four-step analysis. *Step 1:* Remove nodes resected and compute the SC-FC coupling of the remaining network expected after the original resection. *Step 2:* Remove the same number of nodes randomly chosen from the entire network. Compute the overlap between original resection and random resection. Constrain overlap at threshold T (T=1 random resection completely overlaps with original resection, T=0 random resections do not overlap with original resection). *Step 3:* Compute SC-FC coupling of network remaining after random resection (10,000 times) at overlap thresholds from 0.1 to 1 in steps of 0.1. *Step 4:* Compute the z-score of the original resection from the distribution of random resections in each

patient. Z > 0 denotes that the original resections are expected to increase in the SC-FC coupling more so that the resections performed at random. The z-scores with 95% CI are shown in the y-axis of panel c) and the threshold T is shown in the x-axis (star represents $p < 0.05$). Panel c) show that the seizure-free group had significantly increased SC-FC coupling expected post-surgery compared to surgery performed at random. In contrast, the not seizure-free group had decreased SC-FC coupling expected post-surgery compared to random resection, and an alternative seizure may have increased SC-FC coupling. Panel d) expands the data at an example overlap threshold T = 0.3.

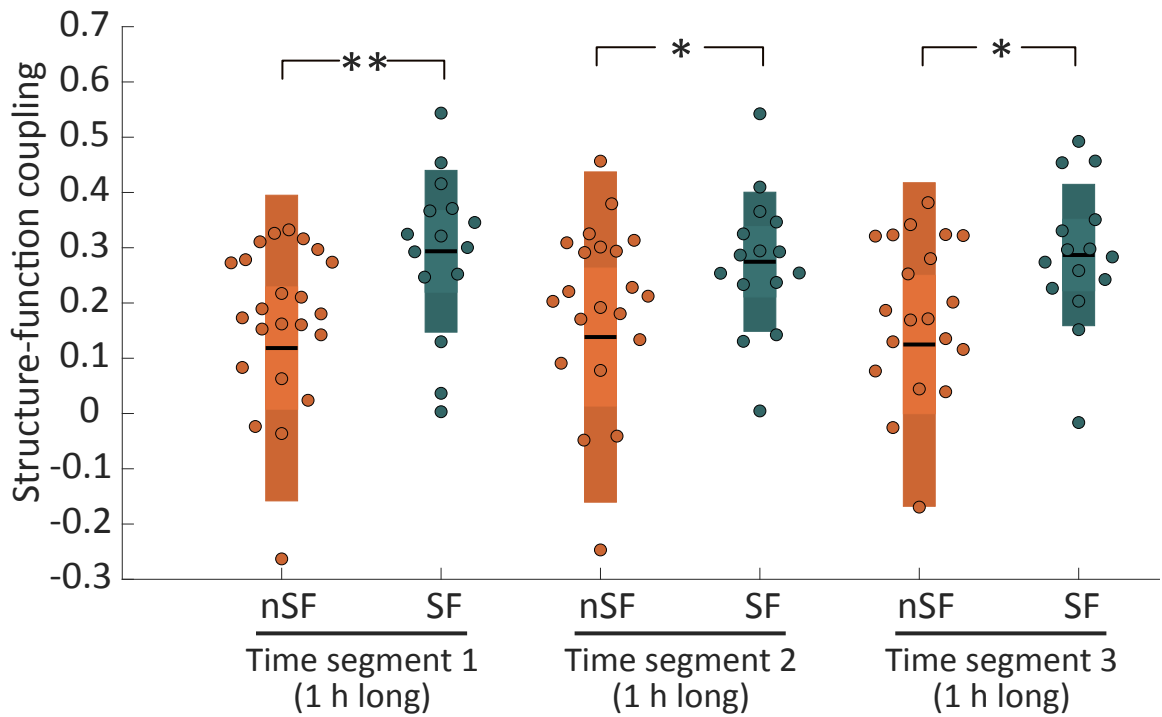

*Figure S5: Consistency of results with expected post-surgery networks estimated from combining structural MRI, diffusion MRI and resection masks.* We estimated the expected post-surgery structural networks following the approach illustrated in Sinha et al. 2021, Neurology[19]; Taylor et al. 2018, Neuroimage Clinical[20]. The box plot shows the structure-function coupling between expected post-surgery structural network and functional networks between iEEG contacts at three interictal segments. Seizure-free individuals had boosted coupling between expected post-surgery structural network and functional networks compared to not seizure-free individuals, regardless of the choice of interictal segments. Statistical estimates—*segment 1: p = 0.003, d = 0.74; segment 2: p = 0.04, d = 0.56; segment 3: p = 0.01, d = 0.68.*

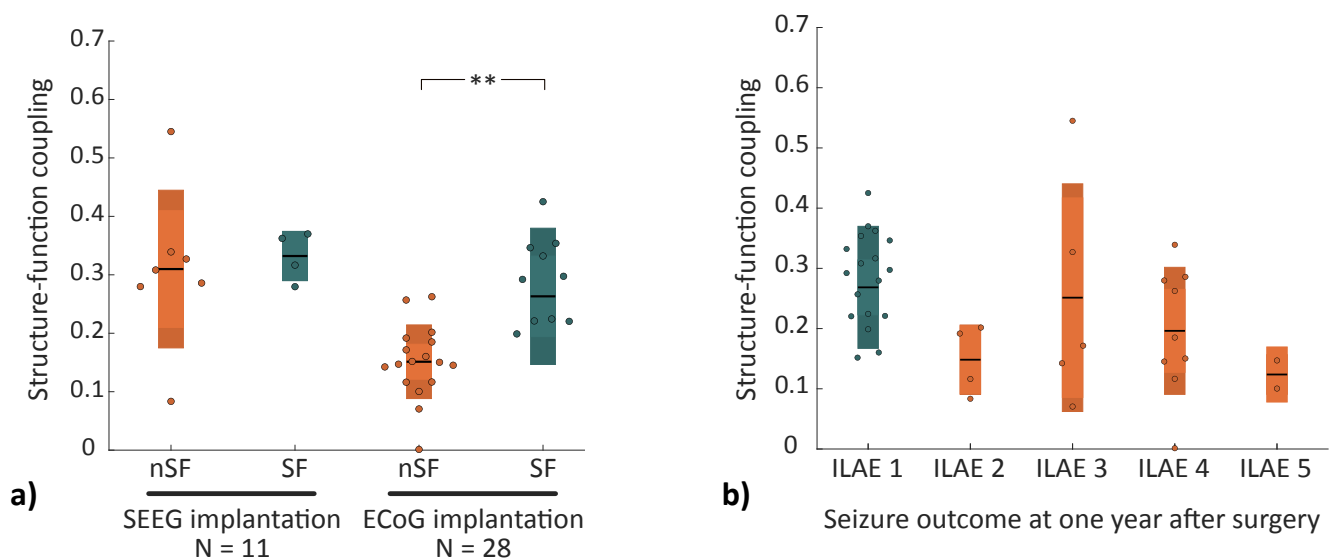

*Figure S6: Effect of implantation and seizure outcomes at one year after surgery.* Panel **a)** groups individuals into seizure-free and not seizure-free categories based on the type of iEEG implantation. 11 individuals were implanted with SEEG, and 28 were implanted with ECoG electrodes. In 28 individuals with ECoG implantations, the seizure-free group had significantly higher structure-function coupling than the not seizure-free group ($p < 0.001$, $d = 1.27$). In individuals with SEEG implantations, the sample size of 11 subjects, 7/4 split, did not reveal any significant effect. As opposed seizure outcome at the last known follow-up reported in the main manuscript, panel **b)** shows the structure-function coupling across different categories of ILAE outcomes at one-year after surgery.

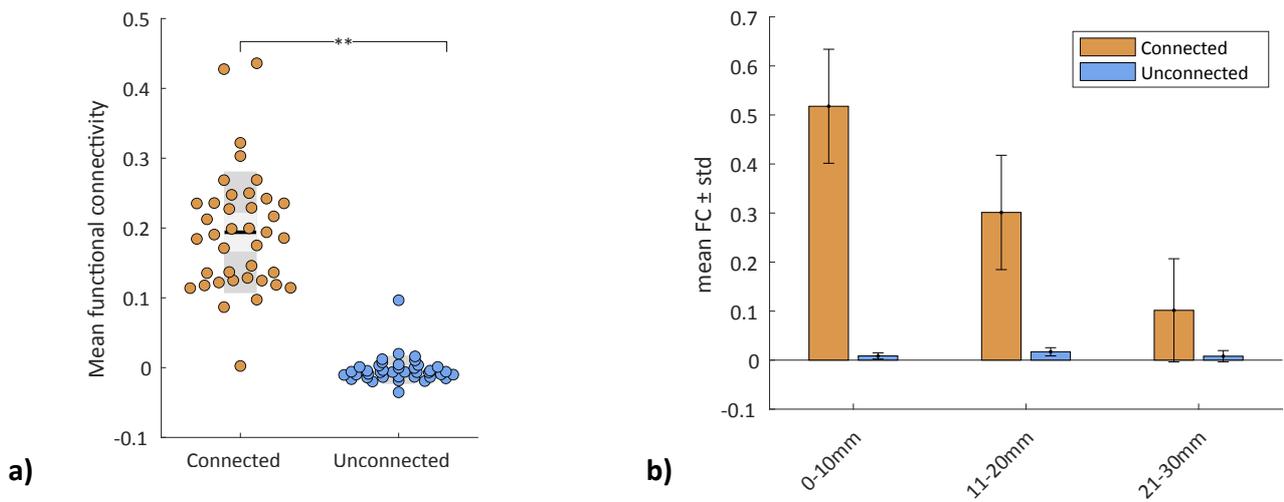

*Figure S7: Relation of FC to connectedness and Euclidean distance.* Two widely reported findings (e.g., Goni et al., 2013, PNAS[45]) reproduced on the SC-FC data generated between iEEG contacts: a) significantly stronger ($p < 0.005$) mean functional connectivity between structurally connected node pairs compared to structurally unconnected node pairs, and b) decline in the strength of functional connectivity between structurally connected and unconnected node pairs in relation to Euclidean distance between contacts.

**Table S1: Demographic and clinical data**

| Subjects | Sex | Side of Surgery | Hippocampal sclerosis | MRI normal | History of FBTCS | History of status epilepticus | iEEG | Surgery Location | Age at surgery (years) | Age at epilepsy onset (years) | Epilepsy duration (years) | Pre-surgery ASMs (n) | # of electrodes | Outcome |
|---|---|---|---|---|---|---|---|---|---|---|---|---|---|---|
| P1 | Female | Right | Yes | No | Yes | No | ECoG | Temporal | 22.2 | 14.0 | 8.2 | 6 | 88 | NSF |
| P2 | Male | Right | Yes | No | Yes | Yes | SEEG | Temporal | 20.6 | 11.0 | 9.6 | 5 | 81 | SF |
| P3 | Female | Left | No | No | Yes | No | ECoG | Temporal | 28.1 | 3.0 | 25.1 | 12 | 52 | SF |
| P4 | Female | Left | Yes | No | Yes | Yes | ECoG | Temporal | 31.0 | 16.0 | 15.0 | 8 | 63 | SF |
| P5 | Male | Left | Yes | No | No | No | ECoG | Temporal | 26.4 | 7.0 | 19.4 | 5 | 26 | SF |
| P6 | Female | Right | No | No | No | Yes | ECoG | Temporal | 30.2 | 2.5 | 27.7 | 9 | 32 | SF |
| P7 | Male | Left | No | No | Yes | No | ECoG | Frontal | 29.6 | 12.0 | 17.6 | 14 | 77 | SF |
| P8 | Female | Right | No | Yes | Yes | No | ECoG | Frontal | 28.2 | 10.0 | 18.2 | 8 | 104 | SF |
| P9 | Male | Left | Yes | No | Yes | No | SEEG | Temporal | 52.9 | 16.0 | 36.9 | 5 | 32 | SF |
| P10 | Male | Left | No | No | No | No | ECoG | Frontal | 27.9 | 14.0 | 13.9 | 5 | 84 | SF |
| P11 | Female | Right | No | No | No | No | SEEG | Temporal | 32.5 | 10.0 | 22.5 | 3 | 54 | SF |
| P12 | Female | Right | No | Yes | Yes | No | ECoG | Temporal | 25.3 | 7.0 | 18.3 | 7 | 102 | SF |
| P13 | Male | Left | No | No | No | No | ECoG | Frontal | 32.5 | 9.0 | 23.5 | 10 | 124 | SF |
| P14 | Female | Left | No | Yes | No | No | ECoG | Frontal | 27.6 | 3.0 | 24.6 | 9 | 109 | SF |
| P15 | Female | Right | No | Yes | Yes | Yes | SEEG | Temporal | 25.0 | 0.7 | 24.3 | 6 | 73 | SF |
| P16 | Male | Right | No | No | Yes | No | ECoG | Temporal | 45.9 | 14.0 | 31.9 | 6 | 71 | SF |
| P17 | Female | Left | No | No | No | No | ECoG | Frontal | 60.3 | 25.0 | 35.3 | 5 | 94 | NSF |
| P18 | Male | Right | No | No | Yes | No | ECoG | Parietal | 32.2 | 12.0 | 20.2 | 4 | 90 | NSF |
| P19 | Male | Right | No | No | No | No | ECoG | Parietal | 28.4 | 7.0 | 21.4 | 7 | 76 | NSF |
| P20 | Male | Right | Yes | No | Yes | No | SEEG | Temporal | 42.7 | 25.0 | 17.7 | 5 | 27 | NSF |
| P21 | Male | Left | No | No | No | No | ECoG | Frontal | 56.3 | 8.0 | 48.3 | 9 | 82 | NSF |
| P22 | Male | Right | No | Yes | No | Yes | SEEG | Frontal | 21.4 | 5.0 | 16.4 | 7 | 64 | NSF |
| P23 | Male | Right | No | No | Yes | No | SEEG | Temporal | 33.5 | 15.0 | 18.5 | 8 | 53 | NSF |
| P24 | Female | Left | No | Yes | Yes | Yes | ECoG | Temporal | 31.6 | 13.0 | 18.6 | 7 | 72 | NSF |
| P25 | Male | Left | No | Yes | Yes | Yes | ECoG | Frontal | 23.3 | 11.0 | 12.3 | 3 | 120 | NSF |
| P26 | Male | Left | No | Yes | Yes | No | ECoG | Temporal | 35.2 | 20.0 | 15.2 | 7 | 58 | NSF |
| P27 | Female | Right | No | Yes | No | No | SEEG | Temporal | 44.7 | 23.0 | 21.7 | 7 | 28 | NSF |

| P28 | Male | Right | No | Yes | No | No | SEEG | Temporal | 47.7 | 14.0 | 33.7 | 15 | 38 | NSF |
| P29 | Male | Left | No | No | Yes | Yes | ECoG | Temporal | 31.7 | 4.5 | 27.2 | 7 | 97 | NSF |
| P30 | Male | Right | No | Yes | Yes | No | ECoG | Frontal | 27.5 | 6.0 | 21.5 | 9 | 122 | NSF |
| P31 | Male | Left | No | No | No | No | ECoG | Parietal | 39.6 | 6.0 | 33.6 | 15 | 51 | NSF |
| P32 | Female | Left | No | Yes | Yes | No | ECoG | Frontal | 23.3 | 8.0 | 15.3 | 6 | 82 | NSF |
| P33 | Female | Left | Yes | No | No | No | ECoG | Temporal | 45.3 | 15.0 | 30.3 | 10 | 57 | NSF |
| P34 | Male | Right | No | Yes | Yes | No | ECoG | Temporal | 26.2 | 13.0 | 13.2 | 4 | 74 | NSF |
| P35 | Male | Left | No | Yes | Yes | Yes | SEEG | Occipital | 22.3 | 1.0 | 21.3 | 9 | 74 | NSF |
| P36 | Female | Right | No | No | Yes | No | SEEG | Occipital | 38.9 | 20.0 | 18.9 | 8 | 50 | NSF |
| P37 | Female | Left | No | No | Yes | No | ECoG | Frontal | 43.0 | 19.0 | 24.0 | 6 | 68 | NSF |
| P38 | Male | Right | No | Yes | Yes | No | ECoG | Temporal | 24.7 | 19.0 | 5.7 | 8 | 72 | NSF |
| P39 | Male | Left | No | Yes | Yes | No | ECoG | Temporal | 26.6 | 15.0 | 11.6 | 2 | 96 | NSF |